\documentclass[aps,amsmath,amssym,amsfonts,twocolumn, showpacs]{revtex4}
\usepackage[dvips]{epsfig}

\newcommand{\xv}{{\bf x}}
\newcommand{\Xv}{{\bf X}}

\newcommand{\uv}{{\bf u}}

\newcommand{\Rv}{{\bf R}}
\newcommand{\tv}{{\bf t}}
\newcommand{\Tv}{{\bf T}}
\newcommand{\Nv}{{\bf N}}

\newcommand{\grad}{{\bf \nabla}}

\begin{document}

\title{Geometry and Optimal Packing of Twisted Columns and Filaments}
\author{Gregory M. Grason}
\affiliation{Department of Polymer Science and Engineering, University of Massachusetts, Amherst, MA 01003, USA}
\begin{abstract}
This review presents recent progress in understanding constraints and consequences of close-packing geometry of filamentous or columnar materials possessing non-trivial textures, focusing in particular on the common motifs of twisted and toroidal structures.  The mathematical framework is presented to relate spacing between line-like, filamentous elements to their backbone orientations, highlighting the explicit connection between the inter-filament {\it metric} properties and the geometry of non-Euclidean surfaces.  The consequences of the hidden connection between packing in twisted filament bundles and packing on positively curved surfaces, like the Thomson problem, are demonstrated for the defect-riddled ground states of physical models of twisted filament bundles.   The connection between the ``ideal" geometry of {\it fibrations} of curved three dimension space, including the Hopf fibration, and the non-Euclidean constraints of filament packing in twisted and toroidal bundles is presented, with a focus on the broader dependence of metric geometry on the simultaneous twisting and folded of multi-filament bundles.
\end{abstract}
\date{\today}

\maketitle
\tableofcontents

\section{Introduction}
Geometrical models of matter have been a cornerstone of physical theories of materials for centuries.  Like Kepler's hypothesis that the emergent symmetries of crystals derive from optimal packings of hard-spherical ``atoms"~\cite{hales}, such models connect the collective physical properties of microscopic particles and molecules to principles of packing for elementary geometrical objects.  From the statistical mechanics of $n$-body clusters in hard-sphere gases and fluids~\cite{mcquarrie}, to properties of granular and amorphous behavior deriving from the so-called random close-packed state~\cite{bernal, liu}, connections between the geometry of sphere packing and many-body behavior of compact, isotropic particles pervades condensed matter. By comparison, the generic principles and emergent behavior of a parallel class of models, what we call {\it filamentous matter}, remains largely unknown.  Filamentous matter refers to assemblies of multiple one-dimensional, or line-like elements, a geometrical motif that appears in diverse materials, formed a range of dimensions spanning nearly seven orders of magnitude in scale~\cite{pan_14}.   Ropes, cables and textiles are familiar examples from macroscopic materials~\cite{costello, hearle}, and physical considerations of the role their structure plays in emergent mechanical properties like tensile strength date back to as least as early as Galileo's work on the strength of materials~\cite{galileo}.  With the advent of modern microscopy came the discovery that rope- and fabric-like assemblies of macromolecular filaments constitute a crucial and broad class of structure elements of biological matter, from the cytoskeleton to extracellular tissue.  
 
 \begin{figure*}
\center \epsfig{file=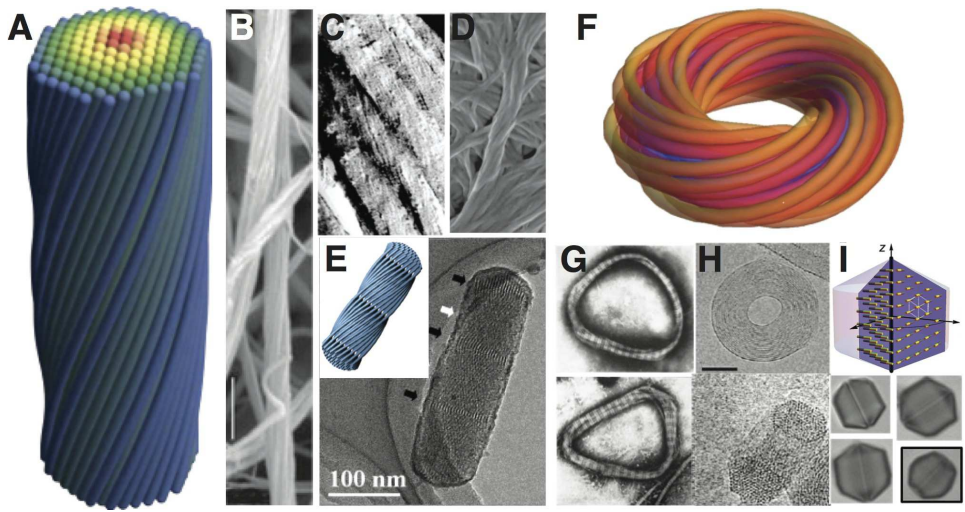, width=5.0in}\caption{ A-E, Twisted bundles (schematic in A) of filaments or columns in biological and synthetic materials (EM images in B-E):  B, fibrin bundles (diameter $\sim 100$ nm)~\cite{weisel}; C,  twisted collagen fibrils derived from tendon (diameter $\sim 100$ nm)~\cite{ottani}; D, twisted fibers of chiral organogel assemblies (diameter $\sim 100$ nm)~\cite{steed}; and E, mesoporous silica templated by twisted columnar assemblies of worm-like surfactant micelles, with schematic in the inset (diameter $\sim 100$ nm)~\cite{yang}.  F-I, Toroidal bundles (schematic of twisted toroidal bundle in F) of filaments or columns from biological and synthetic materials:  G, EM images of twisted toroidal fibers of collagen~\cite{cooper_biochemj}; H, EM images of toroidal condensates of dsDNA~\cite{hud_01}; and I schematic and optical microscopy of faceted columnar droplets of chromonic liquid crystals~\cite{collings_yodh}. }
\label{fig: bundlesandtori}
\end{figure*}

This article reviews recent theoretical advances in the understanding structure formation of cohesive filament assemblies, with the particular focus on how the geometrical interplay between orientation and inter-filament spacing shape the non-trivial structural and thermodynamic properties of assemblies.  Of particular focus will an important class of ``self-twisting" assemblies of filaments or columns, whose complex textures are driven by molecular chirality.   The interplay between chirality and long-range ordering is a subject of long-standing interest in condensed matter, and in liquid crystals in particular~\cite{degennes_prost}.  Inter-molecular forces between chiral molecules favor textures with non-trivial, and twisted, gradients in orientation~\cite{goodby_91, harris_kamien_lubensky}, the simplest example of which being cholesteric order.  Crucially, the patterns of orientation driven by chirality are not always compatible with other types ordering of the given system, as in chiral smectics~\cite{degennes, renn_lubensky, goodby_12}, or even with the geometrical constraints of space itself~\cite{sadoc_frustration} as occurs for the double-twist textures of the liquid crystal blue phases~\cite{sethna_wright_mermin, wright_mermin}.   

The interplay between chiral patterns of orientation and long-range, 2D ordering of columns or filaments, combines these distinct aspects of frustration.  For example, theories of in chiral columnar liquid crystals, show that uniform twist of column backbones and lattice directions, both of which are favored by chirality, is are compatible with bulk columnar ordering~\cite{kamien_nelson_95, kamien_nelson_96}.  Because chiral textures are {\it globally incompatible} with long-range 2D positional order,  in sufficiently chiral bulk systems, twisted textures can only be accommodated through the introduction complex networks of tilt-grain boundaries.  The focus of this review is a related, but distinct, frustration between orientation and 2D positional order that occurs in finite domains with non-trivial textures, in particular, within the twisted structures shown in Fig.~\ref{fig: bundlesandtori}.  Simply put, as one among a broader class of such textures, twist makes it geometrically impossible to evenly space filaments or columns, even {\it locally}, throughout the domain cross-section~\cite{kleman_80, starostin}. 

In filamentous matter, frustration follows from an intrinsic geometric coupling between the orientation and spacing of line-like materials, a relationship which therefore has implications for the structure and thermodynamics of a broad range of self-organized systems.  These include columnar forming liquid crystals, such as lyotropic chromonics (Fig ~\ref{fig: bundlesandtori}I), which exhibit complex and twisted textures upon confinement~\cite{totora_lavrentovich_pnas, collings_yodh}.  When columnar droplets form in a dense chromonic suspensions, the chain-like nature of columns promotes tangential anchoring within droplets, which is known to stabilize toroidal or spontaneously twisted topologies in even achiral chain-like systems~\cite{svensek_podgornik, shin_grason}.  Further examples include twisted and hexagonally-packed worm-like assemblies of chiral (or achiral) surfactant micelles, which have become an important and widely-studied route to chiral mesoporous silica structures (Fig ~\ref{fig: bundlesandtori}E) and provide perhaps the most robust platforms for multi-scale imaging of twisted columnar packing~\cite{che, yang}.  

Beyond columnar systems {\it per se}, cohesive assemblies of two-dimensionally packed filaments or columns constitutes a basic materials architecture in both biological and synthetic systems, relevant to a broader materials context.  In living organisms, assemblies of filamentous proteins represent a primary structure motif, from bundles of cytoskeletal filaments to fibers of extracellular proteins~\cite{alberts}.  Biological filaments are universally helical in structure, owing to the underlying chirality of their constituent macromolecules, proteins and polysaccharides~\cite{bouligand_08, hamley_softmatter_10}.  Hence, rope-like assembles of proteins filaments often exhibit a characteristic to tendency to twist in a handed fashion~\cite{grason_bruinsma_07, grason_09, hagan_prl_10, heussinger}.  The chiral textures of filamentous protein bundles and fibers have been the subject of extensive study in numerous systems, from fibrin~\cite{weisel_pnas_87, weisel} and fibrillar collagen~\cite{cooper_biochemj, bouligand_85, wess, ottani}, to extracellular chitan and cellulose fibers~\cite{neville} to sickle hemoglobin macrofibers~\cite{makowski}.   Beyond structural biofilaments, dsDNA is known to exhibit columnar order at very high concentrations~\cite{livolant_nature}, and also exhibits a both range of chirally-ordered mesophases~\cite{livolant_leforestier} as well as complex topologies, from twisted to folded tori, upon condensation~\cite{hud_95, hud_01} or under confinement~\cite{knobler_gelbart, livolant_pnas_09, livolant_jmb_10}.  Outside of the strictly bioligical contexts, synthetic materials, from peptide-based biomaterial mimics~\cite{rowan} to organogelators and supramolecular polymers~\cite{douglas, meijer}, offer numerous further examples of the self-twisted and densely-packed filament and fibers.

The preponderance of distinct materials exhibiting twisted and densely-packed filaments or columns motivates a series of basic questions regarding the common, underlying geometric principles that constrain their structure.  How does the non-trivial geometry (e.g. twist, bend, etc.) of a columnar assembly influence the structure and energetics of lateral order?  What are the optimal packings of filaments for a given non-trivial assembly geometry, and what factors (geometric, mechanical, molecular) determine these states?  In this review, I discuss recent theoretical progress in understanding  optimal order in twisted columnar and filamentous materials as well as the non-linear interplay between orientation and spacing in columnar systems, more generally.   In particular, this review focuses on understanding how certain patterns of filament orientation critically are {\it incompatible} with homogeneous inter-filament spacing, leading to a frustration of long-range 2D order that is quite analogous to frustration of positional order on intrinsically curved surfaces, like spheres.   The principal goal is to review theoretical frameworks for analyzing constraints of inter-filament spacing deriving from non-uniform textures of two specific types:  twisted, cylindrical bundles (Fig.~\ref{fig: bundlesandtori}A) and twisted toroidal bundles (Fig.~\ref{fig: bundlesandtori}F).  An important focus of the review are models that quantify the thermodynamics costs of non-uniform filament spacing in these {\it incompatible textures}, as well as the nature of the inhomogeneous filament packings that constitute the ground states of these frustrated textures.  

The review is organized as follows.  Sec. \ref{sec: continuum} begins with an introduction to a notion of inter-filament spacing and {\it metric} properties of multi-filament structures in the continuum limit of infinitesimal spacing.  Sec. \ref{sec: twistedbundle} focuses on the unique metric geometry of twisted bundles, relating the constraints of inter-filament packing to those constraining packing on a curved 2D surface, and reviews predictions for the number, type and distribution of defects in the lateral packing of ground-state bundles.  Sec. \ref{sec: tori} reviews theoretical approaches to the structure of twisted toroidal bundles based on ideal properties of filament packings in $S^3$, the three-dimensional hypersphere.  We conclude with a brief discussion of outstanding challenges for understanding optimal packing of filaments and columns beyond the twisted textures considered in this review.  

This review makes extensive use of concepts and methods of classical differential geometry of curved 2D surfaces, principally, the notion of surface metrics and their relation to the intrinsic, or Gaussian curvature.  Though this review relies primarily on graphical descriptions where possible, a reader unaccustomed to these elementary concepts of different geometry may find it useful to refer to an introductory text~\cite{millman} or ``primer"~\cite{kamien_rmp_02} on the subject.

\section{Charting the metric properties of inter-filament packing, a continuum perspective}
\label{sec: continuum}

In this section we illustrate constraints of inter-filament packing deriving from arbitrary, non-uniform textures of filament orientation, and in particular, the connection of these constraints to the metric geometry of 2D curved surfaces~\cite{millman}.  Like membranes or sheets, filaments and columns are extended objects.  Hence, not unlike multi-layered or smectic materials, notions of inter-filament distance are intimately connected to filament orientation.  Even when inter-filament forces are short-ranged, the nature of inter-filament contact is fundamentally non-local.  This is because the relevant ``distance" between a given point, say on one filament, and another filament, say its neighbor, typically refers to the {\it distance of closest approach}, a quantity that depends non-linearly on shape and orientation.

\begin{figure}
\center \epsfig{file=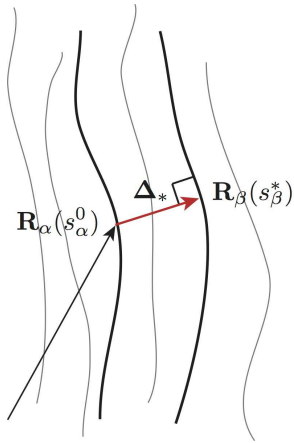, width=1.5in}\caption{ Schematic of local distance of closest approach ${\bf \Delta}_*$ between filament $\alpha$ at $s_\alpha^0$ to filament $\beta$, where $s_\beta^*$ is to closest point to ${\bf R}_\alpha^0$. }
\label{fig: DOCA}
\end{figure}

In collections of filaments, as in condensed phases of multi-filament systems or columnar assemblies, the texture of filament orientations is intrinsically linked to metric (i.e. spacing) properties of inter-filament packing,  quite analogous to the way the geometry, or curvature, of a 2D surface constrains the spacing between material points upon it.  To understand this connection, we consider an ensemble of filaments that are, on average, oriented normal to the $xy$ plane (see Fig. \ref{fig: DOCA}).  Here we focus on the continuum limit, where density is sufficiently high so that filaments are locally parallel  subject to gradual varitation of orientation throughout the packing.  Specifically, we assume that variations of shape and orientation between neighboring filaments are negligible on the scale of inter-filament spacing, set by the diameter $d$.  We consider the spacing between two neighboring filaments, $\alpha$ and $\beta$, whose center lines are described by curves $\Rv_\alpha(s_\alpha)$ and $\Rv_\beta (s_\beta)$, and which intersect the plane at height $z$ at $s^0_\alpha$ and $s^0_\beta$, respectively (see Fig. \ref{fig: DOCA}).  To determine the point of contact from $\Rv_\alpha^0 \equiv \Rv_\alpha(s^0_\alpha)$ to filament $\beta$, we expand the position of filament $\beta$ off of the $z$ plane, $\Rv_\beta (s_\beta) \simeq \Rv_\beta^0 + \Tv \delta s + \kappa \Nv\delta s^2/2$, where $\delta s = s_\beta-s_\beta^0$, and $\Tv$, $\Nv$ and $\kappa$ are the tangent, normal and curvature that approximate the local shape of filament $\beta$ at $z$~\cite{kamien_rmp_02}.  Defining the in-plane separation to be ${\bf \Delta} \equiv \Rv_\beta^0 -\Rv_\alpha^0$,  the square distance between filament $\alpha$ at $z$ nearby points on $\beta$ is simply
\begin{multline}
 |\Rv_\beta (s_\beta)-\Rv_\alpha^0|^2 \simeq | {\bf \Delta}|^2 + 2 \delta s ~( \Tv \cdot {\bf \Delta} )\\ + \delta s^2(1+ \kappa \Nv \cdot {\bf \Delta}) + O(\delta s^3) , 
\end{multline}
which is easily minimized to find the point of nearest contact on $\beta$ at $\delta s_* \simeq - \Tv \cdot {\bf \Delta}/(1+ \kappa \Nv \cdot {\bf \Delta}) $ and the distance of closest approach 
\begin{equation}
\Delta_*^2 =|{\bf \Delta}|^2 - (\Tv \cdot {\bf \Delta} )^2 +O(\Delta^3).
\end{equation}
Hence, the distance of closest approach between nearby filaments is simply the separation measured {\it perpendicular} to the filament orientation.

In the continuum limit we take tangents to be described by a coarse-grained, continuous field $\tv(\xv)$ such that $\Tv_\alpha(s) = \tv \big (\Rv_\alpha(s) \big)$ and consider the square distance of closest approach between infinitesimally spaced filaments, $d {\bf \Delta} = dx~ \hat{x}  +  dy ~ \hat{y} $ ,
\begin{equation}
\label{eq: ds2}
d \Delta^2_*=  g_{ij} (\xv) dx_i dx_j ,
\end{equation}
where $i$ and $j$ sum over in-plane directions and we have defined a {\it metric tensor} to correct for the discrepancy between the distance measured in the plane at $z$ and plane of inter-filament contact
\begin{equation}
\label{eq: gij}
g_{ij} (\xv) = \delta_{ij} - t_i (\xv) t_j (\xv).
\end{equation}
The tensor $g_{ij}$ encodes the intuitive effect that inter-filament spacing may be altered in two ways:  1) either by changing in-plane distance $dx_i^2$; or 2) by tilting filaments along neighbor directions at constant in-plane spacing, reducing true separation.

By drawing on a formal analogy to the metric geometry of 2D surfaces, we may extend our intuition further to understand that certain patterns, or textures, of filament orientation  geometrically frustrate multi-filament packing.  Specifically, we may relate  the constraints imposed by the inter-filament metric, eq. (\ref{eq: gij}), to a {\it dual surface}, $\Xv(x,y)$ carrying the same metric $g_{ij} = \partial_i \Xv \cdot \partial_j \Xv$~\cite{millman}.  Here, duality implies that geodesic distances measured in this surface are equivalent to distance of closest approach between corresponding filaments in the packing, and hence, obstructions to perfect packing of points on $\Xv(x,y)$ imply corresponding obstructions for filament packing at $z$.  

In particular, it is a classical result of differential geometry, well known to cartographers, that the Gaussian curvature of a surface severely constrains distances between objects living upon.    The Gaussian, or intrinsic, curvature $K$ of a surface is simply the product of the two principal curvatures $\kappa_1$ and $\kappa_2$, measured along orthogonal directions for which the lines of curvature are extremal~\cite{millman}.  In general, $K$ may be determined directly from the metric and its derivatives, which has the simple approximate form when the deviation from a flat metric (e.g. $g_{ij} = \delta_{ij}$) is small, $K \simeq -\frac{1}{2} \epsilon_{ik} \epsilon_{j \ell} \partial_k \partial_\ell g_{ij}$ where $\epsilon_{ij}$ is the anti-symmetric tensor~\cite{millman}.  This form is sufficient for analyzing the intrinsic geometry of filament packings where tangents are weakly deflected from the $z$ axis ~\footnote{The small-tilt form for $K_{\rm eff}$ in eq. (\ref{eq: Keff}) is correct to second order in $\tv_\perp$, and is the analog of the small-slope approximation of 2D surface metric in the Monge gauge where surface geometry is described by surface height $h(\xv)$ above the x-y plane for which, $g_{ij} = \delta_{ij} + \partial_i h \partial_j h$.  }.  Defining the {\it effective curvature} $K_{\rm eff}$ of filament packing at $z$ to be the curvature of the dual surface we find
 \begin{multline}
\label{eq: Keff}
K_{\rm eff} \simeq \frac{1}{2} \grad_\perp \times \big [ \tv_\perp ( \grad_\perp \times \tv_\perp) -  ( \tv_\perp \times \grad_\perp) \tv_\perp \big]  \\ = \frac{1}{2} \big[ \partial_{x}^2 (t_y)^2+  \partial_{y}^2 (t_x)^2  - 2 \partial_{x} \partial_{y}  (t_x t_y) \big] ,
\end{multline}
where $\tv_\perp$ is the in-plane filament tilt at $z$ and $\grad_\perp = \hat{x} \partial_{x} +  \hat{y} \partial_{y}$.

\begin{figure}
\center \epsfig{file=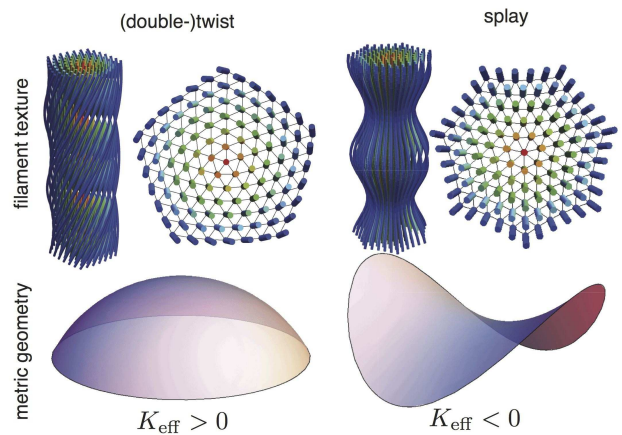, width=3.25in}\caption{Examples of filament textures with positive $K_{\rm eff} >0$ and negative $K_{\rm eff}<0$ effective curvatures, whose equivalent surface geometry is shown schematically with spherical and saddle-like surface patches.  }
\label{fig: effective}
\end{figure}

When $K_{\rm eff} \neq 0$ it is generically impossible for multi-filament systems to maintain uniform spacing throughout the packing, as it is most generally impossible to evenly distribute points on 2D surfaces for which $K \neq 0$~\cite{sadoc_frustration, kleman_advphys_89}.  Hence, the total directive operator $K_{\rm eff}$ plays a special role in the geometry of multi-filament systems, distinguishing textures that are {\it compatible} from those that are {\it incompatible} with uniform inter-filament spacing.  To illustrate the relationship between textures of filament orientation and the dual surface geometry, we consider two characteristic, radially symmetric patterns of in-plane tilt shown in Fig.~\ref{fig: effective}.    A double-twist texture $\tv^{\rm twist}_\perp  = \Omega(y \hat{x} - x \hat{y})$ corresponds to a {\it positive} effective curvature $K^{\rm twist}_{\rm eff}  = 3 \Omega^2 >0$, consistent with a locally spherical geometry of effective radius $(\sqrt{3} \Omega)^{-1}$.  In contrast, for a radial splay texture  $\tv^{\rm splay}_\perp = \gamma (x \hat{x} + y \hat{y})$ we find a {\it negative} intrinsic curvature, $K^{\rm splay}_{\rm eff} = - \gamma^2 < 0$ consistent with a locally hyperbolic, or saddle, geometry with principle radii of curvature $\pm \gamma^{-1}$.  Notice further from eq. (\ref{eq: Keff}) that $K_{\rm eff}$ exhibits a non-trivial dependence on uniaxial vs. biaxial nature of the in-plane texture.  For uniaxial  (cholesteric) twist textures of equivalent pitch the effective curvature is 1/3 value of double twist texture, while $K_{\rm eff} = 0$ for uniaxial (planar) splay.   

\begin{figure}
\center \epsfig{file=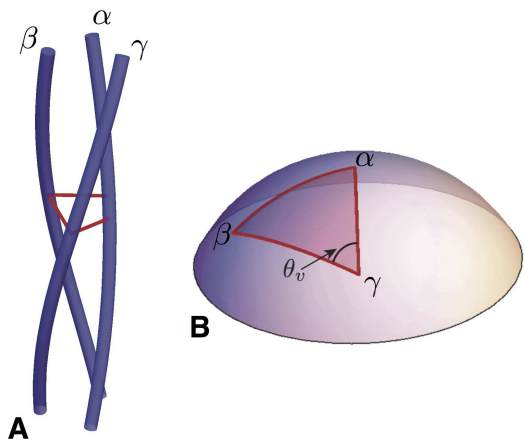, width=2.75in}\caption{In A, a triplet of three twisted filaments, with red lines indicating distances of closest approach between them.  In B, the mapping of inter-filament spacing onto the geodesic separation between three points that form vertices of a geodesic triangle on a positively curved (spherical) surface patch.  The Gauss-Bonnet relates the sum of interior angles (label as $\theta_v$) to the integrated Gaussian curvature within the triangular patch~\cite{kamien_rmp_02}, see eq. (\ref{eq: GB}).  }
\label{fig: GB}
\end{figure}

The implications of ``intrinsically curved"  filament textures, which we deem as {\it incompatible textures}, follow from an application of the famous Gauss-Bonnet theorem~\cite{rubinstein_nelson} relating the Gaussian curvature of a surface to geometry of an equilateral triangle connecting three evenly-spaced points on $\Xv(x,y)$ corresponding to centers of three equally-spaced neighbor filaments in a packing (see Fig. \ref{fig: GB}).  Assuming the geodesic length of each edge is fixed to the preferred inter-filament spacing $d$ the sum of the interior angles $\theta_v$ becomes
\begin{equation}
\label{eq: GB}
\sum_v \theta_v = \pi + \int_{\rm tri} dA ~K_{\rm eff},
\end{equation}
where integral of effective Gaussian curvature is carried out over the dual surface patch enclosed by the triangle, which shows the well known result that the sum of interior angles is greater than (less than) $\pi$ on surfaces of positive (negative) curvature.  Assuming the simplest case for constant $K_{\rm eff}$ within a patch area of $ \Delta A_{\rm tri} $, this formula shows that for close packing, the interior angle between nearest neighbors becomes $\theta_v = \pi/3 + \Delta A_{\rm tri} K_{\rm eff}/3$.  A given filament has $2 \pi$ of surrounding angle available, from which we construct the {\it kissing number} $Z_k$, corresponding to the number of closely-packed filaments which can surround a central filament,
\begin{equation}
Z_k= \frac{6}{1+  \Delta A_{\rm tri} K_{\rm eff}/\pi } .
\end{equation}
Hence, incompatible textures corresponding to positive or negative effective curvature imply $Z_k <6$ and $Z_k>6$, respectively.  In general, for $K_{\rm eff} \neq 0$ the close-packing is incommensurate with integer values of $Z_k$, implying that inter-filament packing {\it must deviate from constant spacing} $d$ and for textures where $K_{\rm eff}\neq0$ inter-filament packing is {\it geometrically frustrated}~\cite{sadoc_frustration}.

The consequence of this geometric frustration is the generation of inter-filament or inter-column stresses for incompatible textures.  A physical model for the energetics incompatible textures is based on the continuum elasticity theory of columnar order~\cite{grason_prl_10, grason_pre_12}, where a free energy functional $F_{\rm col}=\int dV f(u_{ij})$ describes the elastic cost of deformations from a stress-free reference state where filaments/columns are uniformly parallel and possess long-range 2D lattice order transverse to their orientation
\begin{equation}
\label{eq: fel}
f(u_{ij}) =\frac{1}{2} \big[ \lambda (u_{kk})^2 + 2 \mu u_{ij} u_{ij} \big]
\end{equation}
where $u_{ij}$ is the 2D strain tensor describing elastic deformations of the columnar lattice (assumed here to be hexagonal) and $\lambda$ and $\mu$ are the Lam\'e elastic constants parameterizing the cost of deformations of lattice order~\cite{selinger_bruinsma}.  Because columns maintain translational symmetry along their long axis, deformations are described by a two-component displacement field $\uv_\perp(\xv)$, which has components in the 2D plane perpendicular to initial filament orientation, assumed to by the $z$ axis.  Along with 2D positional order, columnar systems possess nematic order associated with the orientations of the columns, $\tv(\xv)$, and transverse displacements deform both types of order.  Column orientations are locked to displacement via 
\begin{equation}
\tv(\xv) = \frac{ \hat{z} + \partial_z \uv_\perp }{\sqrt{ 1+ |\partial_z \uv_\perp|^2}} \simeq \hat{z} + \partial_z \uv_\perp .
\end{equation}
In turn, orientations are coupled to inter-column strains through the strain tensor,
\begin{equation}
u_{ij} \simeq \frac{1}{2} \big( \partial_i u_{\perp j} +  \partial_j u_{\perp i} - t_i t_j\big) ,
\end{equation}
where the geometrically non-linear contribution from in-plane tilt derives from ability of columnar systems to reduce spacing through pure tilt~\cite{grason_pre_12} demonstrated in eq. (\ref{eq: gij}), and therefore, preserves rotational invariance elastic energy (to fourth order in $t_i$).

The {\it inter-column stress} defined by $\sigma_{ij}=df/du_{ij}= \lambda u_{kk} \delta_{ij} + 2 \mu u_{ij}$ is subject to a {\it compatibility condition} which ensures that stresses are compatible with the definition of strain, the geometry of tilt patterns and the topology of displacement field.  The condition derives formally from evaluating anti-symmetric derivatives of strain $\epsilon_{ik} \epsilon_{j \ell} \partial_k \partial_\ell u_{ij}$~\cite{nelson_defects},
\begin{equation}
\label{eq: compat}
Y^{-1} \grad_\perp^2 \sigma_{kk}= s(\xv) - \grad_\perp \times {\bf b}(\xv) - K_{\rm eff},
\end{equation}
where $Y= 4 \mu(\lambda+\mu)/(\lambda+2 \mu)$ is the 2D Young's modulus and $s(\xv)$ and ${\bf b}(\xv)$ are the respective densities of {\it disclinations} and {\it edge dislocations}, respectively, in the transverse lattice order~\footnote{Considering only the elastic energy, the Euler-Lagrange equation for the displacement is $ \partial_j \sigma_{ij} = \partial_z \big[t_j \sigma_{ij}\big]$, which strictly speaking also contributes a term proportional to $\partial_z\big[ \sigma_{ij} \partial_i t_j\big]$ to the right-hand side of (\ref{eq: compat}).} .  This compatibility relation, first derived in ref.~\cite{grason_prl_10}, shows that there are two fundamentally distinct origins of {\it incompatibility} in columnar systems:  topological defects described by multi-valued configurations of $\uv_\perp$ and lattice bond angle; and incompatible orientation textures for which $K_{\rm eff} \neq 0$.  On one hand, topological defects are singular sources of stress, ``quantized" according the discrete symmetries of the underlying 2D lattice, the effective curvature $K_{\rm eff}$ varies continuously, in magnitude and spatial distribution, according the geometry of column orientation.  Accordingly, much like 2D crystalline membranes~\cite{nelson_peliti, nelson_seung}, the effective curvature of columnar/filamentous systems may be viewed as a continuous distribution of {\it disclinations} of local topological charge density $-K_{\rm eff}$~\cite{kleman_advphys_89}.  

In the absence of defects, it is straightforward to determine the energetic costs of incompatible textures.  For example, a bundle of lateral size $R$, eq. (\ref{eq: compat}) implies inter-columnar stresses of order $\sigma \approx Y K_{\rm eff} R^2$ whose energetic cost grows as $F_{\rm col}/V \approx Y (K_{\rm eff} R^2)^2$ implying the elastic costs of geometric frustration are strongly dependent on system size, becoming prohibitive and potentially self-limiting for finite $K_{\rm eff}$ in the thermodynamic limit of $R \to \infty$~\cite{grason_bruinsma_07, grason_09}.  As we show in the next section for twisted bundles, one direct consequence of the geometrically-induced stresses for large $|K_{\rm eff} R^2|$ is the stability of topological defects in the ground state order of incompatible textures.

\section{Topological Defects in Twisted Bundles}
\label{sec: twistedbundle}

\begin{figure}
\center \epsfig{file=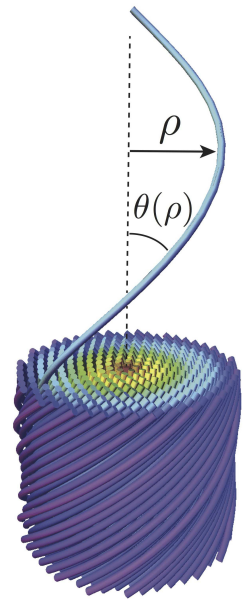, width=1.25in}\caption{A schematic of a (double-)twisted bundle, where color gradient (red to blue) highlights radial distance of filaments from the central filament.  A single, helical filament is shown in the upper portion to highlight the local tilt angle, $\theta(\rho)$, between the filament at radius $\rho$ and the pitch axis. }
\label{fig: twistbundle}
\end{figure}

\begin{figure}
\center \epsfig{file=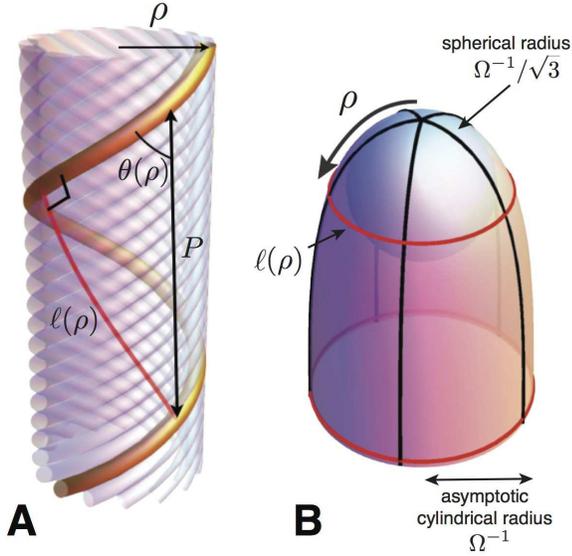, width=3in}\caption{In A, a schematic of the packing of finite diameter filaments at a radial distance $\rho$ from the bundle center.  The amount of space available for packing filaments at $\rho$ is determined by the length $\ell(\rho)$ of a curve (shown in red) between to two points of ``self-contact" along the same filament (shown as orange).  In B, the 2D axisymmetric surface that carries the inter-filament metric properties of a twisted bundle.  The lines of ``latitude" of length $\ell(\rho)$ as defined by the geometry in A.  }
\label{fig: mapping}
\end{figure}

We next consider the optimal structure and energetics of the {\it twisted filament bundle}.  This texture, which for narrow bundles might be recognized as the ``double twist" tube that is the building block of liquid crystal blue phases~\cite{wright_mermin}, is the simplest example of the non-trivial frustration of inter-filament spacing by an incompatible texture.  Here, filament or column backbones are described by the {\it rigid rotation} of in-plane positions about a central axis, say $x=y=0$ along the pitch axis $\hat{z}$.  Filament $\alpha$ in the bundle is described by the helix,
\begin{equation}
\Rv_\alpha (z) = \Rv^0_\alpha + \Rv^0_{\alpha \perp} \big[\cos(\Omega z) - 1\big] + (\hat{z} \times  \Rv^0_{\alpha \perp})\sin(\Omega z)+z \hat{z},
\end{equation}
where $\Rv^0_\alpha$ is the filament position at $z=0$, $\Rv^0_{\alpha \perp}$ is the position in the $xy$ plane at $z=0$ (i.e. vector separation from the central axis) and $2 \pi / \Omega$ is the helical pitch of bundle, which is constant throughout the bundle.  The orientation profile of filaments has the simple form,
\begin{equation}
\tv(\xv) = \cos \theta(\rho) \hat{z} + \sin \theta (\rho) \hat{\phi},
\end{equation}
where the local tilt-angle with respect to the pitch axis follows,
\begin{equation}
\tan \theta (\rho) = \Omega \rho ,
\end{equation}
which goes from $\theta =0$ at the center to $\theta =\pi/2$ as $\rho \to \infty$ indicating an asymptotic approach to circular shape for filaments far from the central axis.  The application of eqs. (\ref{eq: ds2}) and (\ref{eq: gij}) yield the inter-filament metric for a twisted bundle in polar coordinates $(\rho, \phi)$,
\begin{equation}
\label{eq: metric}
d \Delta_*^2 = d\rho^2 + \rho^2 \cos^2 \theta(\rho) d \phi^2 .
\end{equation}
This metric has a simple and familiar interpretation in terms of an axisymmetric dual surface (Fig.~\ref{fig: mapping}), where $\rho$ is the arc-distance from the ``pole" of the surface and $\phi$ is the azimuthal angle around that pole, that is the``longitude"~\cite{bruss_pnas}.  The length of a ``latitude", $\ell(\rho)$, that encircles the pole a distance $\rho$ simply 
\begin{equation}
\ell(\rho) = 2 \pi \rho \cos \theta(\rho) = P \sin \theta(\rho),
\end{equation}
where we used $2 \pi/P = \Omega$.  

In a twisted bundle $\ell(\rho) = 2 \pi \rho/\sqrt{1+ (\Omega \rho)^2}$ can be understood by considering the space available for filaments a radial distance $\rho$ from the center.  The maximum number of filaments that can be placed at $\rho$ is constrained by the length of a curve that passes perpendicular to filaments between two points of contact along the same helical filament (see Fig.~\ref{fig: mapping}A).  In recent studies of closed-packed, $n$-ply geometries~\cite{vanderheijden, olsen_bohr_10}, in which $n$ filament are packed a fixed radius $\rho_p$ from the central twist axis of a ply, the non-linear $\rho$-dependence of $\ell(\rho)$ has been implicated in a surprising ``geometrical jamming" behavior.  The constraints on non-overlap imply a distance between neighbor filaments $d$, a condition which we may approximate at large $n$ by $d \simeq \ell(\rho_p)/n$, and therefore, requires that $\rho_p$ increase with twist as $\rho_p \simeq d /\sqrt{(2 \pi/n)^2-(\Omega d)^2}$.  The filament length per turn of the ply is $L_t(\Omega) =2\pi \Omega^{-1} \sqrt{1+(\Omega \rho_p)^2}$, which when combined combined with divergence of $\rho_p$ at a finite twist ($\Omega \to 2 \pi/(nd)$) implies that the number of turns for fixed-length filaments is a {\it non-monotonic} function of $\Omega$~\cite{olsen_bohr_11}.  That is, $n$-plies achieve a {\it maximum} number of turns at a {\it finite twist} for which $d L_t^{-1}/d \Omega = 0$, a purely geometric phenomenon which we may now relate to the packing of discs on an axisymmetric curved surface.

Given the axisymmetry of the metric in eq. (\ref{eq: metric}), it is straightforward to reconstruct an axisymmetric surface in 3D that encodes the metric properties of the twisted bundle.  Specifically, adopting cylindrical coordinates where $\hat{r}_\perp = \cos \phi \hat{x} + \sin \phi \hat{y}$, the surface has the form,
\begin{equation}
{\bf X} (\rho, \phi) = \frac{ \ell(\rho)}{ 2 \pi } \hat{r}_\perp + z(\rho) \hat{z} ,
\end{equation}
where the function $z(\rho)$ satisfies the  
\begin{equation}
\frac{\partial z}{\partial \rho} =  \pm \sqrt{1 - \cos^6 \theta(\rho)   } ,
\end{equation}
where we used $(2 \pi)^{-1} \partial \ell / \partial \rho = \cos^3 \theta (\rho)$.  This surface is shown in Fig.~\ref{fig: mapping}B, has a tapered, silo-like geometry characterized by the distribution of Gaussian curvature which follows directly from derivatives of the metric~\cite{millman},
\begin{equation}
\label{eq: Keff2}
K_{\rm eff} = -\frac{1}{2 \ell (\rho)} \frac{ \partial^2 \ell(\rho) }{ \partial \rho^2} = 3 \Omega^2 \cos^4 \theta (\rho) .
\end{equation}
This exact form of the curvature distribution agrees with the ``small-tilt" calculation described in the previous section only at the center of the bundle where $K_{\rm eff} (\rho \to 0) = 3 \Omega^2$ and where the geometry of the dual surface is locally well-approximated by sphere of radius $\Omega^{-1} /\sqrt{3}$.  In the large-tilt regime corresponding to points far from the bundle center where $\Omega \rho \gg 1$, the intrinsic curvature vanishes as  $K_{\rm eff} (\rho\gg \Omega^{-1} ) \simeq 3 \Omega^{-2} \rho^{-4}$, indicating an asymptotic approach to a cylindrical geometry for the dual surface.  The concentration of Gaussian curvature at the ``pole" of the dual surface implies frustration of inter-filament packing is largely localized to within a radial distance of order $P$ from the center of the bundle, while sufficiently far from the bundle center, metric constraints permit a nearly regular inter filament spacing, asymptotically commensurate with hexagonal packing, i.e. $Z_k(\rho \to \infty) \to 6$~\cite{bruss_pnas}.  

It is important to recognize that the notion of metric equivalence between twisted bundles and the dual surface is not restricted to the limit of infinitesimally-spaced filaments.  That is, the closest distance between {\it any} two helical curves in the bundle is identical to the {\it geodesic distance} measured between equivalent points on the surface, no matter the separation~\footnote{This follows from the fact that any curve, $C_{12}$, between two points $(\rho_1, \phi_1)$ and $(\rho_2, \phi_2)$ on the dual surface maps onto a unique three dimensional curve, $C_{12}'$, in the bundle which connects filaments at $(\rho_1, \phi_1)$ and $(\rho_2, \phi_2)$ and that intersects all intervening helical curves perpendicular their backbones.  Further, metric equivalence between the surface bundle imply these curves share the identical length (i.e. $L_{C_{12}}= L_{C'_{12}}$).  Likewise, any curve in the bundle maps onto a unique, equal length curve on surface.  Consider the geodesic path $G_{12}$ between two end points on the surface, which maps to curve $G_{12}'$ in the bundle with $L_{G_{12}} =L_{G'_{12}}$.  Because the length of any other curve $C_{12}$ between the same endpoints must have $L_{C_{12}}\geq L_{G_{12}}$, it follows that $G_{12}'$ must also be the shortest possible path between endpoint filaments in the bundle (i.e. a straight line connecting points of contact).}.  This is important because it implies that the duality between the problems of packing in twisted bundles and packing on the dual surfaces holds for finite-sized elements.  For example, we may consider steric, hard tube interactions to prevent inter-filament separations smaller than a diameter $d$.  The duality between packing in bundles and on the dual surface implies that any non-overlapping configurations of (geodesic) discs of diameter $d$ on the surface correspond one-to-one to three-dimensional configurations of non-overlapping filaments of diameter $d$ in the bundle (see e.g. close-packed twisted bundles in ref.~\cite{bruss_pnas}). 

\begin{figure}
\center \epsfig{file=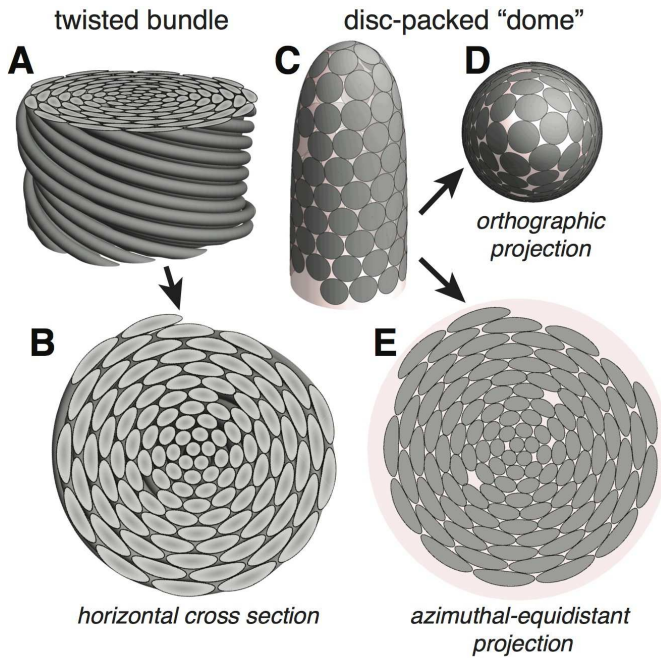, width=3.5in}\caption{Equivalence of finite-diameter filament packing in twisted bundles, and finite-diameter disc packing on ``dome-like" surface carrying the metric of twisted bundle.  The side-view of a twisted bundle is shown in A and the top-view is shown in B, highlighting non-circular shapes of the filament intersections with the plane perpendicular to the pitch axis.  In C, a side view the equivalent disc-packing on the ``dome" shown in Fig. \ref{fig: mapping}B, and two ``polar" projections of the disc-packing are shown in D and E.  The orthographic projection in D, a view from top-down, preserves distances along the azimuthal direction while compressing distances along the radial direction.  In the azimuthal equidistant projection in E, preserves radial distances while stretching azithumal distances, producing the identical image of the filament intersections (azimuthally stretched discs) shown in B. }
\label{fig: disc_map}
\end{figure}

The equivalence between discrete packings of finite diameter elements provides a useful way to illustrate and understand the metric equivalence between bundles and their dual surfaces.  Consider a horizontal section of a twisted bundle as shown in Fig.~\ref{fig: disc_map}A-B and note the apparent ``warping" of the circular cross-sections of the helical tubes in the sections:  horizontal ``slices" of filaments near the bundle center remain circular due to the normal interaction with horizontal plane, while slices towards the outer edge of the bundle stretch, or warp, azimuthally due to the increased tilt.  Consider also the equivalent disc packing on the dual surface shown in Fig.~\ref{fig: disc_map}C.  Due to the non-zero Gaussian curvature of the dual surface, any projection of the disc packing to a planar surface will distort the image of the disc packing with a local geometry that varies throughout the projected image, familiar from contential distortions in cartographic projections of from the globe~\cite{snyder}.  See, for example, the disc packing in orthographic projection (i.e. ``viewed from above" in Fig.~\ref{fig: disc_map}D), where discs appear compressed along the radial directions away from the pole at the center of the image.  Viewed from another projection which maintains distances measured along the radial direction (see Fig.~\ref{fig: disc_map}E)  known as the {\it azimuthal equidistant} projection,  we find that projected image of dual surface disc packing is identical to images of the planar section of bundle normal to the pitch.  Other planar sections of the bundle, those not necessarily normal to the pitch axis, correspond to {\it azimuthal equidistant} projections of the same disc packing where the center of the image is no longer a point of axial symmetry of the dual surface (the pole at $\rho =0$).  The warping of filament sections in the planar cut of a bundle has long been recognized as in the context of the so-called ``contact" problem in textiles and yarns~\cite{hearle, pan_02}, though only recently has the connection to non-Euclidean geometry been understood.

\subsection{Disclinations in Twisted Bundles}
\begin{figure}
\center \epsfig{file=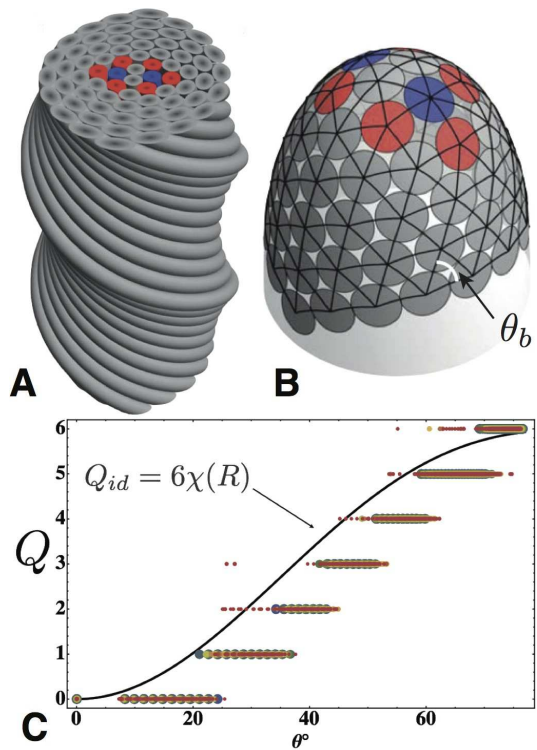, width=2.75in}\caption{A simulated ground state of an $N=70$ twisted bundle from ref. \cite{bruss_pnas} is shown in side-view in A and along with the corresponding disc-packing on the bundle-equivalent surface in B.  Triangulation of the packing on the curved surface yields the nearest-neighbor ``bond network", identifying defects in the packing as deviations from six-fold coordination of the bond network (i.e. disclinations).  While filaments with six neighbors are colored gray, five- and seven-fold coordinated filaments are shown as red and blue, respectively.  In C, the total topological charge of the ground-state packing $Q$, defined in eq. (\ref{eq: Q}), plotted as a function of the twist angle of the outermost filament in the bundle, $\theta=\arctan (\Omega R)$, with the colored data points showing results from simulated ground states and the solid line showing the geometric prediction for the ``ideal" topological charge given by eq. (\ref{eq: Qid}). }
\label{fig: mapping2nd}
\end{figure}

The non-Euclidean metric geometry and the associated global and local constraints on inter-filament packing implied by the mapping have critical consequences for physical models of cohesive filament assembly in twisted bundles.   The Gauss-Bonnet theorem and its application to triangulations of disc packings on the dual surface may be exploited to derive the relationship between bundle twist, the topology of the nearest-neighbor bond network in the bundle and the deformation of ideal inter-filament geometry\cite{bruss_pnas},.  Figure \ref{fig: mapping2nd} shows a filament bundle and its dual representation as a curved-surface disc packing.  Because the geodesic distances measured on the surface represent the true inter-filament spacing, the triangulated network of nearest-neighbor bonds on the surface properly encodes the topology of nearest inter-filament contact.  In particular, from the triangulation of the dual packing we may count the neighbor statistics of filaments in the packing, and its deviation from the 6-fold packing of a parallel bundle.  Denoting  the number of filaments (or discs) in the bulk of the bundle (not a surface vertex) possessing $n$ neighbor bonds by $V_n$, we define the total {\it topological charge} of the bundle to be
\begin{equation}
\label{eq: Q}
Q= \sum_n (6-n)V_n .
\end{equation}
This definition is consistent with the definition of topological {\it disinclination} charge where points of 5-fold (7-fold) coordination in the bond network correspond to +1 (-1) contributions to $Q$~\cite{nelson_defects}.  Applying eq. (\ref{eq: GB}) by summing over the triangulated faces of nearest-neighbor mesh and using the facts that 1) each face is spanned by three edges (or ``bonds") 2) each edge connects two vertices and 3) each internal (non-surface) vertex accounts for $2 \pi$ total internal angle we arrive at a generalized version of the Euler-Poincar\'e formula~\cite{kamien_rmp_02}
\begin{equation}
\label{eq: euler}
Q - 6 \chi = N_b \langle \delta \theta_b \rangle ,
\end{equation}
where 
\begin{equation}
\chi= \frac{1}{2 \pi} \int_{\rm mesh} dA ~K_G ,
\end{equation}
quantifies the total integrated Gaussian curvature within the triangulated packing and the right-hand describes the $N_b$ internal angles of boundary vertices, $\theta_b$, from equilateral packing with
\begin{equation}
 \langle \delta \theta_b \rangle = \frac{1}{N_b} \sum_b (\theta_b - \pi/3) .
\end{equation} 
where $\theta_b$ is shown schematically in Fig.~\ref{fig: mapping2nd}B.

Typical applications of the Euler-Poincar\'e formula consider triangulations without boundary (say, for crystalline packings on surfaces of spherical topology), such that the right-hand side is zero ($N_b =0$)and $Q$ is a topological invariant, fixed by the Euler characteristic $\chi$~\cite{giomi_bowick_advphys}.  In the case of a twisted bundle, the total {\it disinclination} charge is not a topological invariant.  The deficit between $Q$ and $6 \chi$ will be accommodated by packing deformation at the boundary (i.e. $\langle \delta \theta_b \rangle \neq 0$).  Nevertheless, eq. (\ref{eq: euler}) provides a useful heuristic for understanding the structure of low-energy packings by noting that the right-hand side a measure of the inter-filament strain in the packing.  Intuitively, one expects that interactions that favor equidistance filaments will favor equilateral packing at the boundary (specifically in the limit of $d\to 0$ where area per face vanishes), hence, $\langle \delta \theta_b \rangle \neq 0$ indicates a locally sub-optimal geometry.   More specifically, the magnitude of inter-filament strain, or the variation of inter filament spacing, implied by $\langle \delta \theta_b \rangle \neq 0$ can be understood in terms of the mean geodesic curvature $\kappa_g \approx 3\langle \delta \theta_b \rangle/d$  of lattice row in a nearly triangular packing of average spacing $d$.    Due to the row curvature, the change in spacing between successive rows is roughly $ \kappa_g d^2$.  For a bundle with a  number of radial rows $N_r$, the relative change of spacing between filaments at the center and periphery of the bundle, respectively $d_0$ and $d_b$, becomes $d_b/d_0 -1  \approx N_r \langle \delta \theta_b \rangle$.  For 2D bundles where $N_r \propto N_b$, it follows from eq. (\ref{eq: euler}) that $Q - 6 \chi$ is indeed proportional to the excess separation between filaments at the bundle surface relative to the center.

\begin{figure*}
\center \epsfig{file=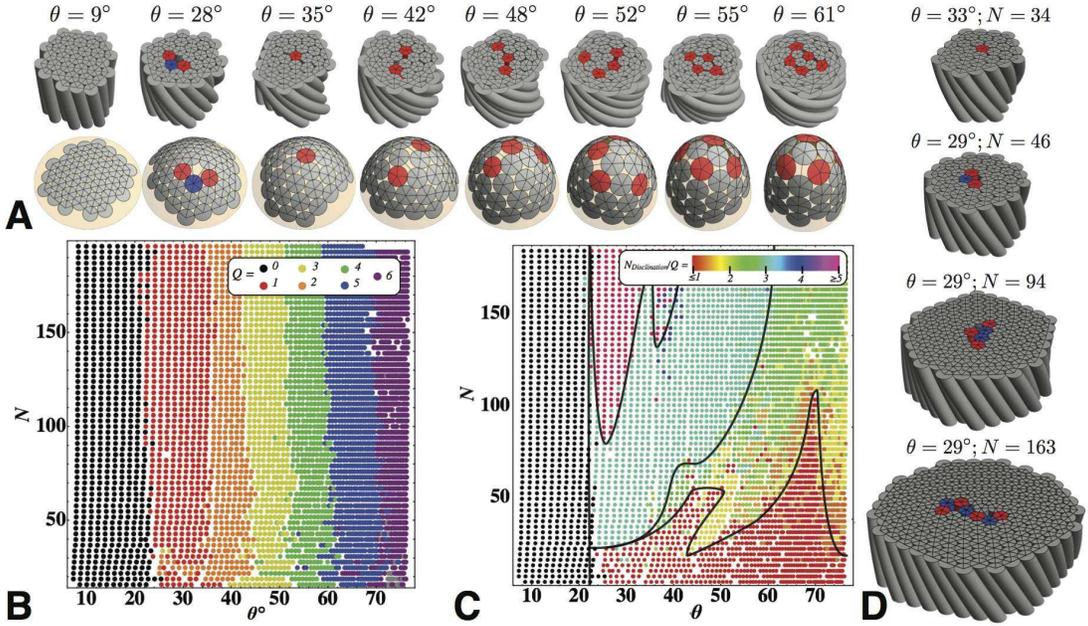, width=5.75in}\caption{Simulated ground states of twisted filament bundles adapted from refs. \cite{bruss_pnas, bruss_softmatter}.  In A, optimal packings of a 34-filament bundle, with increasing twist angle showing an increase in the number of five-fold (disinclination) defects, shown in red.   The total topological charge of simulated ground states shown in B for bundles of variable twist angle, $\theta=\arctan(\Omega R),$ and filament number $N$.  In C, the number of disclinations per charge, $N_{Disclination}/Q$, is shown for simulated ground states, with dark lines drawn to guide the eye to regions of roughly constant value.  In D, a series of simulated ground states at fixed $\theta \approx 30^\circ$ (corresponding to $Q=1$) with increasing $N$, showing the transition from compact disclinations to extended ``charged scars" of alternating 5-7 defect pairs.}
\label{fig: phase}
\end{figure*}

As cohesive interactions favor uniform inter-filament spacing throughout, a simple conjecture is that in {\it energy-minimizing bundles} the packing prefers values of topological charge where $\langle \delta \theta_b \rangle = 0$, such that the {\it ideal topological charge} may be defined as $Q_{id} \equiv 6 \chi$.  Assuming that bundle cross sections retain a roughly circular shape, we may calculate the dependence of $Q_{id}$ on twist and radius of bundles,
\begin{equation}
\label{eq: Qid}
Q_{id} = \frac{3}{\pi} \int ~d \rho ~ \ell(\rho) K_{\rm eff} (\rho) = 6 \big[ 1- \cos ^3 \theta (R)  \big] ,
\end{equation}
where we use $dA = d \rho~ \ell(\rho)$ and eq. (\ref{eq: Keff2}).  This simple relationship makes three significant predictions about the optimal (energy-minimizing) packing of twisted bundles.  First, the preferred disclination charge of bundles is {\it independent} of filament diameter, depending only on the tilt angle $\theta$ at the surface of the bundle, which itself is fully determined by the ratio $R/P$.  Second, for $\theta \neq 0$, $Q_{id}\geq 0$, indicating a preference for excess 5-fold coordinated ($Q=+1$) sites in the bundle cross section.  Third, the preferred topological charge of the packing increases from $Q_{id}= 0$ at $\theta =0$ to a maximum of $Q_{id}= 6$ as $\theta \to \pi/2$.

\begin{figure}
\center \epsfig{file=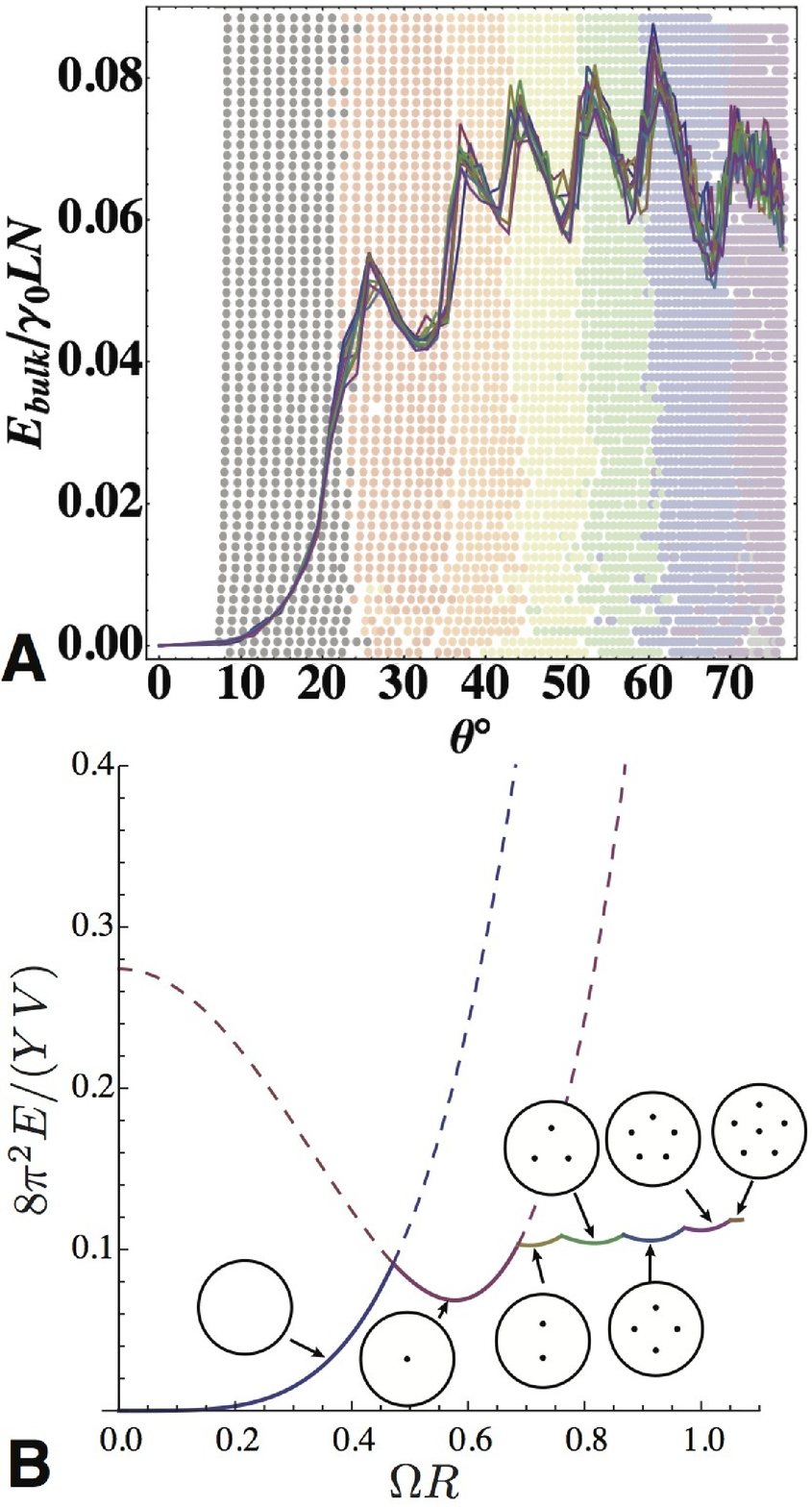, width=2.5in}\caption{The ``bulk" energy density (total energy minus excess energy of surface filaments) of simulated ground states of twisted bundles vs. twist angle, for large bundle sizes $N=166-193$ as computed in ref. \cite{bruss_softmatter} is shown in A.  Energy curves are overlaying the results for $Q$, highlighting the coincidence of multiple minima in the energy density with step-wise transitions in optimal value of $Q$.   The shape of the simulated bulk energy density is compared to continuum elasticity theory calculations for twisted bundles possessing only five-fold disclinations, calculated in refs. \cite{grason_prl_10, grason_pre_12}, is shown in B, with optimal arrangement of defects shown for each distinct branches (corresponding to distinct $Q$ values) of energy minimal.  The dashed lines show the metastable branches of defect free and $Q=1$ elastic energy density, which meet at the transition point $(\Omega R)_c = \sqrt{2/9}$ (corresponding to $\theta_c \simeq 25^\circ$). }
\label{fig: evstwist}
\end{figure}

These predictions for the optimal distribution of defects in the cross section of twisted filament bundles have been tested in the context of numerical simulations of cohesive filament bundles~\cite{bruss_pnas, bruss_softmatter}.  These simulations employ a simple stochastic algorithm to optimize the cohesive energy of an $N$-filament bundle with fixed twist $\Omega$.   Here, the finite-diameter $d$ of filaments enters as the energy-minimum of pair-wise cohesive interactions, which was assumed to have a form similar to an Leonard-Jones potential in which the separation is the distance of  closest approach between helical centerlines of filaments.  Fig. \ref{fig: mapping2nd}C compares the $Q_{id}$ to the topological charge $Q$ of numerically-minimized bundle packings for $N=16 - 196$, which is extracted directly from triangulated neighbor packing that has been conformally mapped the plane.  Notwithstanding its continuous $\theta$-dependence as well as the simple assumption of cylindrical bundle shape, the form of $Q_{id}$ in eq. (\ref{eq: Qid}) does a remarkable job to capture the increase in the excess of 5-fold defects of numerical ground-state packings.  

As shown in Fig. \ref{fig: phase}B which maps the minimal-energy value of $Q$ in terms of $\theta$ and $N$, these simulations confirm that the net topological charge is largely determined by twist angle (or equivalently by the integrated curvature on the dual surface) and independent of filament number.  The evidently universal dependence of $Q$ on $\theta$ is all the more surprising when analyzing the dependence of other structural measures of the packing on $\theta$ and $N$.  For example, in Fig. \ref{fig: phase}C we show total number of disclinations per topological charge $Q$ (where ``disclination" here refers to any non 6-fold coordinated filament in the bulk packing), which unlike $Q$ itself, exhibits a complex and non-universal dependence on both filament number and bundle twist.  One critical observation is the abundance of excess $5-7$ pairs in the energy-minimizing states of large-$N$ bundles (Fig.~\ref{fig: phase}D), a trend which is not-unlike the formation of ``grain boundary scars" on spherical~\cite{bowick_travesset_nelson, bausch_science} and catenary~\cite{irvine_nature} surfaces at large-$N$.

Despite these obvious complexities in detailed ground-state structure (both in numbers, positions and charge of individual defects), optimal bundles maintain a fixed and universal value of {\it net} number of 5-fold defects as measured by $Q$ for a given $\theta$ consistent with the purely geometric considerations implied by the dual-surface mapping.  The universal evolution of $Q$ with twist implies a corresponding universality in the $\theta$-dependence of the energy of the bundle.  Fig. \ref{fig: evstwist}A shows the plots of $E_{bulk}/V$ ``bulk" energy density (total - surface filament energy) vs. $\theta$ for simulated ground states in the range of $N$.  Again, despite the complex differences in detailed packing structure, for large-$N$ the bulk energy shows a characteristic dependence on $\theta$ that is dominated in the underlying and universal changes in $Q$.  At low angle, the energy of a defect-free ($Q=0$) bundle exhibits a roughly power-law increase with $\theta$.  The monotonic $\theta$-dependence holds until a critical value of $\theta_c \simeq 25^\circ$, at which point the ground-state becomes unstable to an excess 5-fold defect, $Q=1$, marked by a cusp and secondary minimum, indicating the mitigating effects of defects in highly-twisted bundles.   Further cusps appear that the transitions to higher integer $Q$, leading a characteristic ``saw-tooth" dependence of $E_{bulk}/V$ on $\theta$ in the defect-mediated regime.  Notably, an energetic landscape of similar structure was calculated in the context of continuum elasticity theory calculations of twisted bundles~\cite{grason_prl_10, grason_pre_12} possessing energy-minimizing configurations of 5-fold disclinations (Fig.~\ref{fig: evstwist}B) .  At small twist, predictions of the continuum theory appear quantitatively consistent for small twist (notably, continuum theory predicts a critical angle of $\theta_c = \arctan( \sqrt{2/9})\simeq 25.2^\circ$ in good agreement with simulations). It should be noted that the small-tilt approximation underlying this theory lead to qualitative failures at large-twist, including an unbounded increase in $Q$ as $\theta \to \pi/2$.

\subsection{Dislocations in Large-$N$ Bundles}

Five-fold disclinations are evidently favorable in sufficiently twisted bundles, yet these topologically ``charged" defects are not the only means of relaxing geometrical frustration in bundles.  Indeed, Azadi and Grason showed that for sufficiently large bundles ($N\gg1$) excess disclinations which appear only above a critical threshold of twist $\theta_c \simeq 25^\circ$, are pre-empted by a class of topologically neutral defects that become stable at lower twist~\cite{azadi_grason}.  These defects, edge dislocations in the cross-section, are ``bound" 5-7 pairs~\cite{nelson_defects}, which correspond to a partial row of filament positions that terminates within the bulk of the packing.  Because these defects are only energetically stable at sufficiently large $N$, dislocation-only ground states of twisted bundles have not been characterized via the numerical methods applied for stable disinclination patterns for $N \lesssim 100$.  Nonetheless, the regime of large bundles size $R/d \gg 1$ and low-twist $\theta \ll1$ where multi-dislocation patterns emerge as minimal-energy configurations is well suited to the continuum elastic theory of 2D ordered bundles outlined in the previous section.

\begin{figure}
\center \epsfig{file=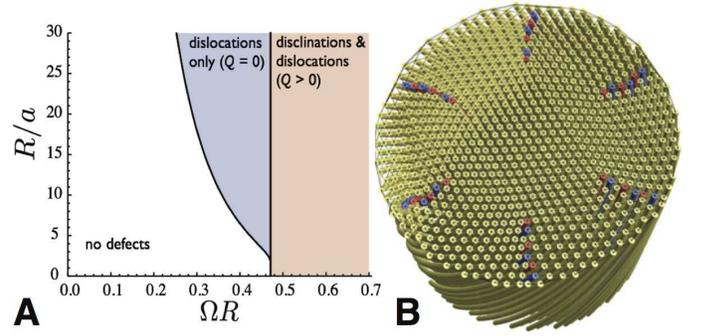, width=3.5in}\caption{In A,  stability phase diagram for defects in twisted bundles, calculated from continuum elasticity theory~\cite{azadi_grason}, showing regions where dislocations (neutral 5-7 disclination pairs) and ``charged" defect configurations possessing at least one excess five-fold disinclination are stable relative to the defect free bundle.  Here, $a$ is the inter-filament lattice spacing.  In B, schematic of the ``scarred", multi-dislocation ground state at intermediate twist for sufficiently large bundles (i.e. $R/a\gg1$), with five- and seven-fold coordinated filaments shown in red and blue, respectively.    }
\label{fig: dislocations1}
\end{figure}

The stability of dislocations can be understood by considering the stress distribution in a defect-free twisted bundle and the work done to remove a partial row of filament positions in the bundle, to create an edge dislocation~\cite{azadi_grason}.  The dominant contribution to the stress derives from the tilt-induced azimuthal compression of inter-filament spacing at the outer periphery of the bundle, from which we can crudely estimate the magnitude of this stress as $\sigma_{\phi \phi} \approx - Y t_\phi^2 = - Y (\Omega \rho)^2$.  A more careful calculation shows that the stress profile of the defect-free state $\sigma_{\phi \phi} = 3 Y \Omega^2/128 (R^2-3 \rho^2)$ is only compressive sufficiently far from the bundle core ($\rho \geq R/\sqrt{3}$)~\cite{grason_pre_12}.  To maximize the energy relaxation upon introducing a dislocation, we may consider a Volterra construction~\cite{chaikin_lubensky}, in which dislocations correspond to the removal a material along a cut in the bundle cross section.  Due to the compressive stress at the bundle periphery, stable dislocations have polarizations corresponding to Burgers vectors locally aligned to azimuthal direction and the removal of a partial row of filament positions extending radially from the dislocation (at $\rho \lesssim R$) to the free edge of the bundle.   Following standard arguments~\cite{peach_koehler}, removing a row of filament positions of width $b \simeq d$ and of length $\ell \approx R$ corresponds to a relaxation of the elastic energy by roughly $\sigma_{\phi \phi} d R$, from which we estimate the  energy of twist-dislocation coupling $E_{\rm twist}$ to be
\begin{equation}
E_{\rm twist} \approx - Y b \Omega^2 R^3 .
\end{equation}
Comparing this to the elastic ``self-energy" of introducing a single dislocation in the cross section $E_{\rm disc} \approx Y b^2 \ln (R/b)$~\cite{chaikin_lubensky} we estimate the critical degree of bundle twist at which dislocations become stable,
\begin{equation}
(\Omega R)_{\rm disl}^2 \approx \frac{b}{R} \ln(R/b) .
\end{equation}
Significantly, while the stability condition of isolated {\it disclinations} is predicted to be independent of bundle size (i.e. $(\Omega R)_{\rm disc} =\sqrt{2/9}$) the threshold twist for appropriately polarized {\it dislocations} is 1) highly dependent on $R/d$ and 2) found to decrease with increasing bundle size, vanishing in the $R/b \to \infty$ limit.  Notably, an essentially equivalent argument was first developed in the context of ``neutral" dislocation patterns formed in 2D crystalline assemblies on curved surfaces with open boundaries by Vitelli, Irvine and Chaikin, yielding a similar increase in dislocation stability as the ratio of crystal size to lattice spacing grows~\cite{irvine_nature}.



A more careful analysis of the position dependence of the elastic energy of dislocations in twisted bundles yields the defect stability diagram shown Fig. \ref{fig: dislocations1}A.  For sufficiently, narrow bundles $R/b \lesssim 3$ the dislocations and disclinations are predicted to become energetically preferably at roughly the same degree of large twist, comparable to $(\Omega R)_{\rm disc} =\sqrt{2/9}$.  In contrast, for mesoscopically large bundles where $R/b \gg 1$  dislocations become stable in twisted bundles at degrees of twist far below the threshold for excess 5-fold disclinations, predicting a broad range of multi-dislocation ground states at intermediate twist for large bundles.  To put this into context, we may compare these thresholds with the observed size and twist angles of self-twisting filament assemblies.  For example, fibrin bundles~\cite{weisel_pnas_87}  and twisted collagen fibrils~\cite{wess} are observed to have twisted angles in the ranges of $8^\circ-10^\circ$ and  $15^\circ-17^\circ$  respectively, which are both well below the threshold angle for stabilization of a single 5-fold disinclination $\theta_{\rm disc} \simeq 25^\circ$.  For comparison, for bundles of mesoscopic dimensions typical for fibrin and collagen $R \approx 100 d$, the elastic theory predicts that dislocations become favorable above a threshold twist of $\theta_{\rm disc}(R/b=100) \simeq 9^\circ$, below or comparable to the observed twists of either structure.  These observations suggest that while excess disinclinations may not be stable in of some of the most commonly observed twisted filament architectures, stable dislocations and multi-dislocation patterns are likely features of optimal packing of these materials.

\begin{figure}
\center \epsfig{file=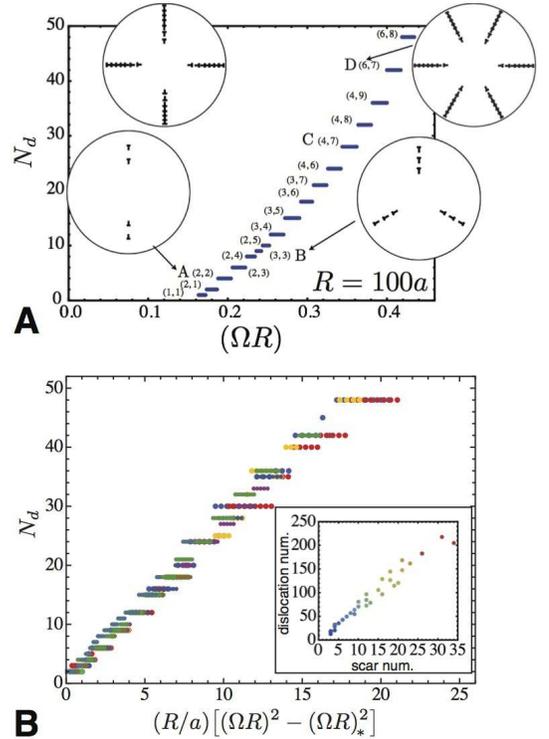, width=2.75in}\caption{In A, optimal number of dislocations vs. bundle twist in a $R=100a$ bundle, from continuum theory of twisted bundles~\cite{azadi_grason}.   The integer pair $(m,n)$ refers to structures with $m$ ``scars" each possessing $n$ dislocations.  B shows, the collapse of total number of defects  (from bundles $R/a=20-700$) with parameter $R/a[(\Omega R)^2 - (\Omega R)_*^2]$ where $  (\Omega R)_*$ is the critical twist for stable dislocations.  The inset of B, shows the proportionality between the total dislocation number and optimal scar number, predicting roughly 6 dislocations per scar independent of $R/a$.  Here, the color scale indicates the gradient in bundle sizes, with red and blue, corresponding to large and smaller $R/a$, respectively.  }
\label{fig: dislocations2}
\end{figure}

The structure and thermodynamics of multi-dislocation ground states of twisted bundles was studied using the Greens functions for dislocation sources of stress in cylindrical bundles to calculate the elastic energy of defect competing patterns~\cite{azadi_grason} . For bundle twist in excess of the critical dislocation twist $(\Omega R)_{\rm disl}$, the energetically preferred number of dislocations follows a characteristic scaling with bundle twist and size (see Fig.~\ref{fig: dislocations2}A).   This scaling can be understood in largely geometric terms by balancing the length of azimuthal compression at the free boundary on the dual surface $|\ell(R) - 2 \pi R |\approx R (\Omega R)^2$ with the azimuthal length $N_d b$ removed by $N_d$ radial lattice rows of width $b$ removed from the periphery of the bundle, yielding
\begin{equation}
N_d \sim\frac{R}{b} (\Omega R)^2  .
\end{equation}
The optimal symmetries of multi-dislocation patterns have also been explored in the context of ground states of twisted bundles, and more recently, the context of the dual problem of crystalline ``caps" on spherical surfaces~\cite{davidovitch_grason_pnas, azadi_grason_prl}.  For $N_d\gg 1$, minimal energy patterns of dislocations are shown to be radial chains of dislocations, or ``neutral scars", extending from the free edge and terminating the bulk of a bundle (see Fig.~\ref{fig: dislocations1}B).  This motif of a ``neutral" 5-7 disinclination chain,   originally dubbed ``pleats" in based on observations of colloidal assemblies on curved 2D surfaces~\cite{irvine_nature}, has the structure along its length of a tilt grain boundary separating two orientationally mismatched regions by an angle $\delta \phi \simeq b/D$, where $D$ is the spacing between dislocations along the scar.  While ordinary grain boundary do not terminate in the bulk of the crystal, the ``tips" of scars do, and therefore, act as singular, disclination-like points around which the lattice orientation rotates rapidly by $\delta \phi$.  It was recently shown~\cite{azadi_grason_prl} that the an elastic competition between these distinct portions of scars --- on one hand the ``line tension" of the scars which prefers to localize dislocations into a small number of high-angle grain boundaries and on the other hand the disclination-like tips of scars which favor alternatively a larger number of small-angle grain boundaries --- select an optimal number of scars $n_s\sim N_d$ which diverges in direct proportion to the number of dislocations as $R/b \to \infty$.  Fig.~\ref{fig: dislocations2}B  shows the linear relationship between $N_d$ and $n_s$ for simulated ground state patterns of dislocations of bundle sizes in the range of $R/b=20-700$.  Remarkably, these results predict that the ratio$N_d/n_s$, the number of dislocations per scar, approaches a universal value ($\approx 6$ from the slope of $N_d$ vs. $n_s$ in Fig.~\ref{fig: dislocations2}B), independent of lattice spacing, bundle twist or other materials parameter in the asymptotic limit $R/b \to \infty$.

\section{Twisted Tori in Curved (and Flat) Space}

\label{sec: tori}

In this section we review geometrical approaches to the problem of twisted filament packing based on studies of {\it fibrations} of the three-sphere ($S^3$).  As was first understood in the context curved-space models of the blue phase chiral liquid crystals, the ambient positive curvature of $S^3$ admits uniform double-twist textures which are otherwise frustrated in Euclidean space ($R^3$)~\cite{sethna_wright_mermin}.  This fact provides a means to construct and study ``ideal" twisted structures in curved space whose structure only becomes heterogeneous, perhaps defect riddled, upon projection to $R^3$.  

A second important feature of the twisted fibrations of $S^3$, particularly their projections to $R^3$, is that they provide a natural means to construct twisted toroidal bundles.  Like the straight bundles of the previous section, in twisted toroidal bundles filament positions rotate around a central backbone along its contour, but unlike straight bundles, twisted toroids have backbones that also bend around into a closed curve.  Toroidal assemblies of filaments and columns are known to form in a variety of systems, from condensed biopolymers like DNA~\cite{hud_95, hud_01, livolant_pnas_09} or collagen~\cite{cooper_biochemj} to columnar droplets of chromonic liquid crystals~\cite{collings_yodh} , and therefore, a generic model of structure and thermodynamics of inter-filament packing in this geometry has broad value.  Beyond its potential application to any of these materials systems, the physical and geometry theory of packing in twisted toroidal bundles provides a natural way to analyze the interplay between bundle geometry and inter-filament organization, beyond straight, twisted bundles.  Simply put, how are the metric properties and consequences thereof altered when a filament bundle is twisted and simultaneously {\it bent}?

In the context of liquid crystalline materials, the unique geometry of textures in $S^3$ first drew interest as a conceptual approach to ``defrustrating" double-twist textures which are characteristic of chiral, blue phases~\cite{wright_mermin}. Kl\'eman was first to consider how metric properties of ideal fibrations -- that is, properties beyond orientation  --  would be relevant to physical models of twisted filament packing, albeit, filaments embedded in an unphysically curved space~\cite{kleman_85}.   More recently, Sadoc and Charvolin have expanded on this initial analysis by exploring a more general class of fibrations and their projections to twisted toroidal bundles in $R^3$~\cite{sadoc_charvolin_epj, sadoc_charvolin_jphys}.  In this section, we aim to provide primarily a descriptive summary of the key properties of twisted filament packing geometry in $S^3$, metric features of their projections to $R^3$ and the connection to the twisted, straight bundle packing problem of the previous Sec.~\ref{sec: twistedbundle}.  An interested reader will find considerably more detailed analyses of three-sphere fibrations in \cite{sadoc_charvolin_jphys}.

\subsection{Double-twisted filament packings in $S^3$}

$S^3$ can be constructed as a three-dimensional submanifold of a four-dimensional (Euclidean) space satisfying
\begin{equation}
\label{eq: S3}
x_1^2+x_2^2+x_3^2+x_4^2 = \Omega^{-2} ,
\end{equation}
where $\Omega^{-1}$ is the radius of the 3-sphere, which we will see can be related to twist of embedded filament packings.  Critical to models of filament packing is the structure of fibrations of $S^3$~\cite{sadoc_frustration}, which are decompositions of this space into a collection of non-intersecting curves, or fibers, such that every point maps on to a unique curve.  Like the case of the straight bundles in $R^3$ above, the fibrations of interest here are also equipped with an important property that every fiber is associated with a unique point on a lower dimensional manifold (a 2D surface), such that distance between fibers in $S^3$ (i.e. the distance of closest approach) is encoded in metric properties of the surface, known as a {\it base}.  

\begin{figure}
\center \epsfig{file=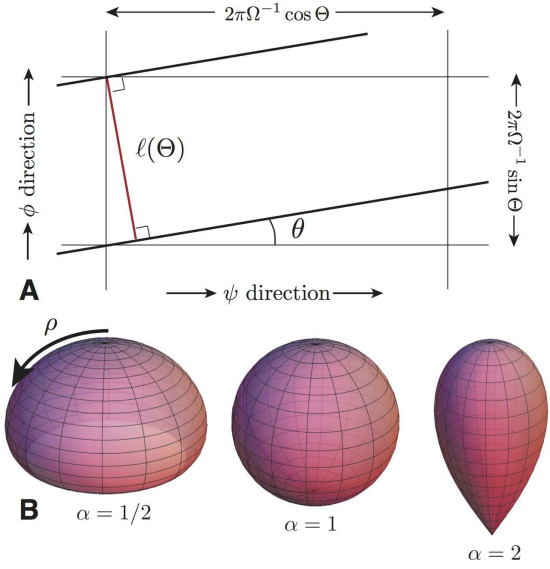, width=3.25in}\caption{In A, The toroidal coordinate system of fibrations in $S^3$, showing the dimensions of the $T^2$ unit cell at fixed $\Theta$.  Fibers (filament backbones) wind along the dark solid lines at an angle $\theta$ with respect to the $\hat{\psi}$ axis.  The perimeter is defined as the distance of closest contact between the fiber at $\phi$ and its periodic image at $\phi+2 \pi$.  In B, surfaces (bases) carrying the inter-filament metric of fibrations in $S^3$.  }
\label{fig: Hopftorus}
\end{figure}

The topological and metric properties of the 3-sphere are encoded in the following toroidal coordinates,
\begin{eqnarray}
\nonumber
x_1 &=& \Omega^{-1} \cos \phi \sin \Theta \\ \nonumber
x_2 &=& \Omega^{-1} \sin \phi \sin \Theta \\ \nonumber
x_3 &=& \Omega^{-1} \cos \psi \cos \Theta \\ \nonumber
x_4 &=& \Omega^{-1} \sin \psi \cos \Theta ,
\end{eqnarray}
a parametrization that satisfies eq. (\ref{eq: S3}) by construction.  Surfaces of fixed $\Theta$ are periodic under $\phi \to \phi + 2 \pi$ and $\psi \to \psi + 2 \pi$ and therefore have the topology of 2D tori.  In these coordinates the metric of $S^3$ has a simple form,
\begin{equation}
dx_i^2= \Omega^{-2} \big(\sin ^2 \Theta   d \phi^2 +  \cos^2 \Theta d \psi^2 + d \Theta^2 \big) \ \ {\rm in} \ S^3 ,
\end{equation}
which shows that the metric of fixed $\Theta$ surfaces is Euclidean, and spans a rectilinear periodic cell of dimensions $2 \pi \Omega^{-1} \sin \Theta$ and $2 \pi \Omega^{-1} \cos \Theta$, along the $\phi$ and $\psi$ directions respectively (see Fig.~\ref{fig: Hopftorus}A).  Fibers, or filament backbones, are curves running along surfaces of constant $\Theta$ parameterized by
\begin{equation}
\phi (\psi)= \phi_0 + \alpha \psi .
\end{equation}
Here, $\psi$ plays the role of an arc coordinate, describing different positions along the filament backbone, and $\alpha$ is the number turns of the fiber around the $\phi$ direction per rotation around the $\psi$ direction (see Fig.~\ref{fig: Hopftorus}A).  It is straightforward to show that any two such curves sharing the same $\alpha$ (at different $\phi_0$) remain {\it equidistant} along their entire length.  

Like the case of the straight bundles in $R^3$ the inter-filament metric may be deduced by considering the length, $\ell(\Theta)$, of a curve separating points of ``self-contact" along a given fiber in $S^3$ which defines the ``perimeter", or the amount of space available for packing fibers along the $\phi$ direction at fixed $\Theta$~\footnote{This notion of ``perimeter" neglects self-contact with any periodic images of the fiber that may pass between $\phi_0$ and $\phi_0+ 2 \pi$.  For example, for $\alpha = n/m$ (where $n$ and $m$ are relatively prime integers) a fiber will wind $n$ times around the $\phi$ direction for every $m$ turns round the $\psi$, leading to $n$ copies of the fiber section between $\phi_0$ and $\phi_0+ 2 \pi$, which clearly limit the number of filaments of a given diameter that can be packed on the 2-torus.  Because the implicit $n$-fold symmetry of the fibration around $\phi$ has been neglected at this stage, some care must be taken when applying this result to packings, specifically, every filament in the packing must be associated with the $n-1$ copies spaced at intervals of $2 \pi/n$.   In their analysis of metric properties of $S^3$, Charvolin and Sadoc retain the $n$-fold images of a fiber when constructing the $\ell(\Theta)$ leading to an somewhat modified formula for the base metric. }.  As shown in Fig.~\ref{fig: Hopftorus}, this is easily reduced to $\ell(\Theta) =  2 \pi \Omega^{-1} \cos \theta \sin \Theta$ where $ \theta = \arctan \big( \alpha \tan \Theta\big)$ is the ``tilt-angle" of the fiber on the 2-torus at $\Theta$.  Hence,
\begin{equation}
\ell (\Theta) =2 \pi \Omega^{-1}\frac{  \sin \Theta \cos \Theta}{ \sqrt{  \cos^2 \Theta + \alpha^2 \sin^2 \Theta} } 
\end{equation}
Using this perimeter and noting the distance between fibers at different $\Theta$ is simply $\Omega^{-1} |d \Theta |$, we have the inter-fiber metric in $S^3$,
\begin{equation}
\label{eq: metricS3}
d \Delta_*^2 (\alpha) =(2 \Omega)^{-2}\Big[ d ( 2 \Theta)^2 + \frac{  \sin^2 (2 \Theta)d \phi_0^2 }{\cos^2 \Theta + \alpha^2 \sin^2 \Theta } \Big] \ \ \ \ {\rm in} \ S^3.
\end{equation}
This metric formula shows that geometry of the ``base" surface for twisted fibrations of $S^3$, like that of the case of straight, twisted bundles in $R^3$, is axi-symmetric, with $\Omega^{-1} \Theta$ the arc-distance from a ``pole" at $\Theta=0$.  Again, the independence of the inter-filament metric on $s$ derives from the equidistance of any pair of curves sharing the same $\alpha$.  

\begin{figure}
\center \epsfig{file=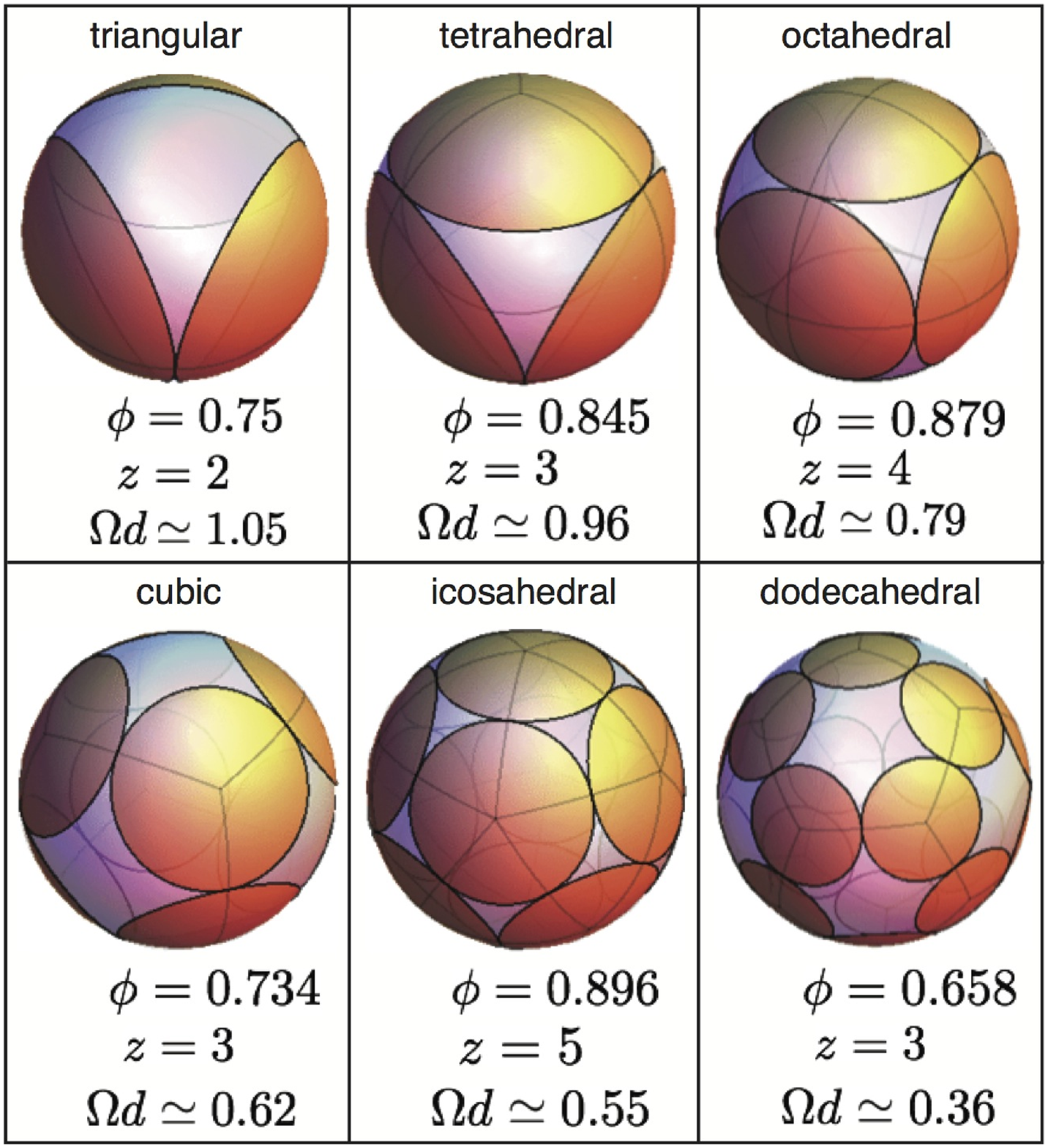, width=3in}\caption{The ``ideal" packings of twisted and equally-spaced filaments (diameter $d$) in $S^3$ whose positions correspond to vertices of Platonic solids projected on $S^2$, where $\phi$, $\Omega d$, and $z$ denote packing fraction, reduced twist and coordination number of the packing~\cite{kleman_85}.   }
\label{fig: hopfpack}
\end{figure}

The particular case of $\alpha =1$ corresponds to the celebrated {\it Hopf fibration}~\cite{sadoc_frustration}, where each fiber is a closed geodesic of $R^3$ (great circle) which winds (twists) around its neighbor once every cycle from $\psi$ to $\psi+ 2 \pi$.  The inter-fiber metric has the remarkably simple form, $d \Delta_*^2 (\alpha=1) = (2 \Omega)^{-2}\Big[d ( 2 \Theta)^2 + \sin^2 (2 \Theta)  d \phi_0^2 \Big]$, identical to the geodesic distance measured between points on a 2-sphere of radius $(2 \Omega)^{-1}$, with polar and azimuthal angles $2 \Theta$ and $\phi_0$, respectively.   In this unique geometry ($\alpha =1$), packing double-twisted filaments in $S^3$, maps {\it identically} onto the generalized Thomson problem of packing points on $S^2$~\cite{saff, altschuler}.  Exploiting the homogeneous metric geometry of the Hopf fibration, Kl\'eman constructed a class of ``ideal" twisted filament packings in $S^3$, such that all nearest neighbor filaments are closely-packed, at a center-to-center spacing equal to the diameter, $d$~\cite{kleman_85}.  Evenly spaced distributions of discs on $S^2$ are only possible for certain numbers of discs, or equivalently, for certain ratios of diameter to sphere radius, $2 \Omega d$, packings which correspond to the vertices of the Platonic solids which possess only a small number of discs ($\leq20$).  By mapping the Hopf packings in $S^3$ to their associated  ``Platonic" packing on $S^2$ (shown in Fig.~\ref{fig: hopfpack}) Kl\'eman showed that the densest such packing of twisted filaments has icosahedral symmetry with each filament surrounded by five neighbors~\cite{kleman_85}.  

The cases of $\alpha \neq 1$ provide generalizations of the Hopf fibration, known as {\it Siefert fibrations}~\cite{sadoc_charvolin_jphys}.  When $\alpha \neq 1$, the fibers are not geodesics of $S^3$, though they remain closed curves for any rational $\alpha$.  Examples of the 2D base (embedded in $R^3$) are shown in Fig.~\ref{fig: Hopftorus}B, where we take $\rho = \Omega^{-1} \Theta$ to be the arc distance from the pole at $\Theta =0$, and define the Euclidean distance from the $z$ axis of radial symmetry to be,
\begin{equation}
r_\perp(\rho) = \frac{\sin( 2\Omega \rho) }{2 \Omega \sqrt{1+(\alpha^2-1) \sin^2(\Omega \rho) } } .
\end{equation}
From this metric, it straightforward to show the Gaussian curvature at the pole has the following form,
\begin{equation}
\label{eq: KGS3}
K_G(\rho=0)= \Omega^2 (1 + 3 \alpha^2) \ \ \ \ {\rm in} \ S^3.
\end{equation}
which is consistent with the apparent increase of curvature with $\alpha$ seen in Fig.~\ref{fig: Hopftorus}B~\footnote{Eq. (\ref{eq: KGS3}) also provides a direct illustration of {\it O'Neill's theorem}~\cite{berger}, which equates the Gaussian curvature of the base surface of a fibration to the sum ambient curvature of the embedding space (here $\Omega^{-2}$) and three times the twist of the fibration (here $(\alpha \Omega)^2$).  Notice that the same formula holds for the straight twisted bundle in Euclidean space (zero ambient curvature).}.  Note that for $\alpha \neq 1$, the 2D surface is not smoothly embeddable in $R^3$.  For $\alpha < 1$, surface cannot be extended beyond a cusp at $\rho_m$ where $|\partial_\rho r_\perp( \rho_m)| = 1$, while for $\alpha>1$ conical singularity develops at the $\Theta = \pi/2$ (or $\rho = \Omega^{-1} \pi/2$) pole. To date, optimal packings geometries on these base metrics of Seifert fibrations for $\alpha \neq 1$ have not been studied.

\subsection{Projecting ``Ideal" Packings to Euclidean Space}

In principle, the high symmetry of the inter-fiber distances of the fibrations of $S^3$ (all fibers are equidistant and metrics are axisymmetric) provides a natural setting for investigating optimal packings of twisted filaments with a more complex topology than the straight, twisted bundle described in Sec.~\ref{sec: twistedbundle}.  However, exploiting the ideal properties of fibrations of $S^3$ in models of filament packing in Euclidean space requires overcoming at least two critical challenges.  First, for a given filament number, $\alpha$, $\Omega$ and model of filament interactions, the optimal filament packing must be identified, which is the analog of the generalized Thomson problem defined for the broader class of base surfaces.  Provided these optimized packings can be determined for $S^3$, an additional step is needed to ``rescue" the filament configurations from curved space ($S^3$), via some projection to $R^3$, which in turn alters the inter-filament distances form their ``ideal" geometry in $R^3$.  As it is not possible to project from curved space to a flat while globally preserving distance properties, one might view the choice of projection from $S^3$ to $R^3$ as a second, and currently unsolved, step of the optimization procedure.

\begin{figure}
\center \epsfig{file=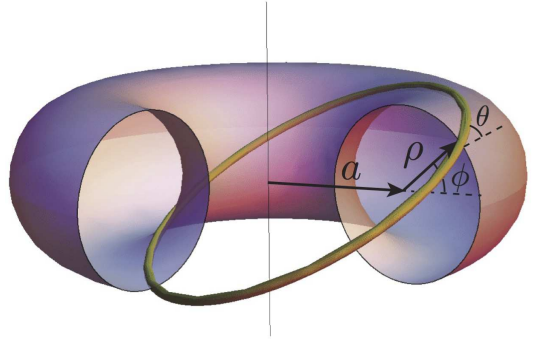, width=2.75in}\caption{A schematic of toroidal coordinates under stereographic projection to $R^3$.  Surfaces of constant $\Phi$ are concentric tori (pink), and fibers/filaments (gold) wind along these surfaces around both the minor and major axes of the tori, where $\theta$ is the (constant) angle between tangents and the circular axis of the torus.  The circular fiber shown here corresponds to $\alpha =1$, a projection of the Hopf fibration.}
\label{fig: stereotorus}
\end{figure}

\begin{figure*}
\center \epsfig{file=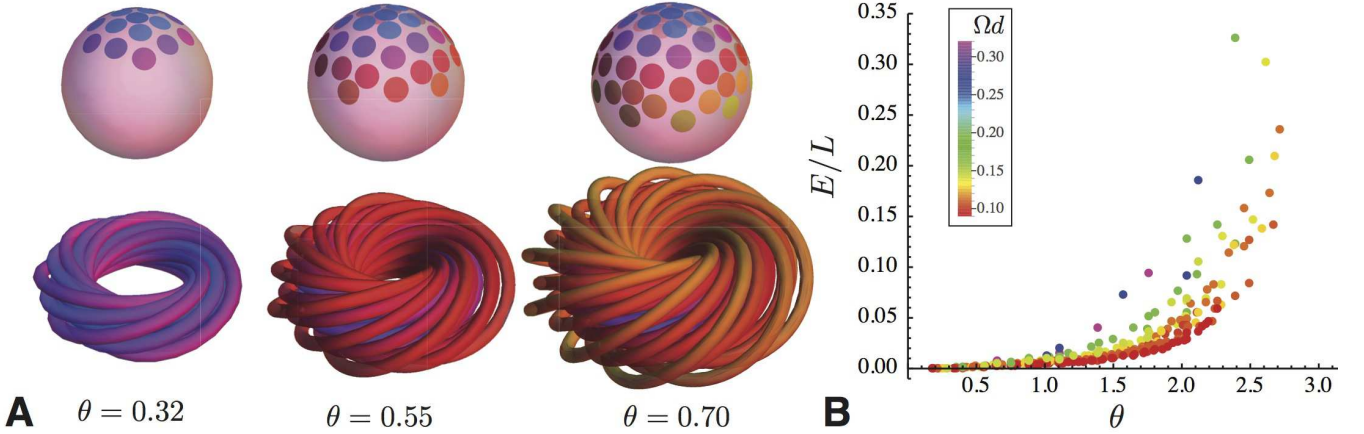, width=7in}\caption{In A, stereographic projections of twisted filaments packings in $S^3$ to $R^3$, based on the Hopf vibration ($\alpha = 1$).  Filament positions on the $S^2$ base derive from the icosodeltahedral tesselations, where the value of $\Omega d$ is chosen based on distance between the central filament (at the pole in $S^2$) and its first shell of neighbors.  In A, $\{2,2\}$ packings are shown, both on $S^2$ (top) and projections to $R^3$ (bottom).  From left to right shows examples with increasingly larger maximum $\Phi$, corresponding to larger polar distance on $S^2$, larger toroidal thickness, larger twist angles of outer most filaments.    In B, the strain energy density defined by eq. (\ref{eq: stereoE}) and calculated numerically and plotted versus twist angle of outer filaments for icosodeltahedral tesselations, for a range of tesselations from small twist, $\Omega_{\{4,2\}} = 0.0895 d^{-1}$, to large twist, $\Omega_{\{1,1\}} = 0.325 d^{-1}$.} 
\label{fig: stereotori}
\end{figure*}

One approach that has been suggested by Sadoc and Charvolin is based the stereographic projection from $S^3$ to $R^3$~\cite{sadoc_charvolin_epj}.  Because it is conformal, the stereographic projection has the advantage of preserving angular properties, including the skew angle of neighboring filaments in the double-twisted packing.  Furthermore, the metric distortion from the optimal geometry of $S^3$ vanishes near the projected ``pole" of the stereographic image, such that appropriate choices of the projection pole allow different (finite) regions of the $S^3$ packing to be projected to $R^3$ with nominal distortion of inter-filament spacing.  For example, a projection that generates toroidal bundles that preserve the curved-space metrics along their center line takes $x_1$ as the projection axis so that filament positions in Euclidean coordinates become
\begin{equation}
x(s) = \frac{x_3(s)}{1 - \Omega x_1(s)}; y(s) = \frac{x_4(s)}{1 - \Omega x_1(s)}; z(s) = \frac{x_2(s)}{1 - \Omega x_1(s)} .
\end{equation}
Under this projection, the filaments wind around a family of nested tori.  The central axis of the torus is the $\hat{z}$ axis and a toroidal coordinates --- $a$ is the ``major radius", or the distance of torus center from the axis and $\rho$ is the ``minor radius", or the radial distance of the torus surface from the torus center (as in Fig.~\ref{fig: stereotorus}) --- related to $S^3$ coordinates by,
\begin{equation}
\rho(\Theta) = \Omega^{-1} \tan \Theta; \ a(\Theta) = \Omega^{-1} \sec \Theta,
\end{equation}
Hence, the pole at $\Theta =0$ maps to the planar circle of radius $\Omega^{-1}$, and the pole at $\Theta = \pi/2$ maps to the central ($z$) axis (infinite radius circle).  As the stereographic projection is conformal, the angle $\theta$ between the axial direction and filament tangents winding around fixed-$\Theta$ tori remains constant, $\tan \theta = \alpha \tan \Theta$.  The interfilament metric in of stereographic projection has the form,
\begin{equation}
\label{eq: metricR3}
d \Delta_*^2 (\alpha) =\frac{\omega^2}{(2 \Omega)^{2}}\Big[d ( 2 \Theta)^2 + \frac{  \sin^2 (2 \Theta)d \phi_0^2 }{\cos^2 \Theta + \alpha^2 \sin^2 \Theta } \Big] \ \ \ \ {\rm in} \ R^3,
\end{equation}
which is identical to metric in $S^3$, eq. (\ref{eq: metricS3}), up to the conformal factor representing a locally-isotropic scaling of dimensions:
\begin{equation}
\label{eq: omega}
\omega = \frac{1}{1 - \cos( \phi_0 + \alpha \psi) \sin \Theta } .
\end{equation}
For filament packings, this conformal factor represents the failure of the projected fibrations to maintain equidistance (as $\psi$ advances) along their length.  Filament positions on the inner-(outer-)side of the torus correspond to $\cos( \phi_0 + \alpha \psi) >0$ ( $<0$), and hence $\omega>1$ ($<1$) describes the measure of over- (under-)crowding in toroidal packing, an affect which increases in magnitude for tori of large minor radius, or larger $\Theta$ (see, for example, projections in Fig.~\ref{fig: stereotori}A).  Geometrically, the variation of inter-filament spacing can be understood in terms of the difference between inner and outer spacing between consecutive, non-concentric toroidal layers, as well as non-uniform angular rotation of filament positions around the central axis of the tori required to maintain constant $\theta$ around a torus.  

As a consequence of the conformal distortion of inter-filament spacing, the optimality of ``ideal" packings in $S^3$ when projected stereographically to $R^3$ becomes compromised, increasingly so as $\Theta$ increases.  In particular, it is unclear at which point distortions of inter-filament spacing become sufficiently large that the ``ideal" packings identified in $S^3$ fail to provide accurate model, even at a qualitative level, of the constraints and energetic consequence of packing in twisted toroidal bundles.  

Absent a projection from curved space that preserves the equidistance of fibers in $R^3$, one can nevertheless, consider the energetic costs of inter-filament strains as a measure of the excess frustration cost of {\it bending} a twisted bundle into a torus.   Setting aside the extent to which this excess cost could be relaxed by local or global adjustments of filament position and orientation in packing, we illustrate this cost for the class of toroidal bundles projected from the Hopf fibration ($\alpha = 1$).  

The stereographic projection of the Hopf fibration has the feature that filament trajectories, which are (great) circles in $S^3$, are mapped to circles to in $R^3$ (see e.g. filament in Fig.~{fig: stereotorus}) .  This fact and the formula for the closest distance from point to a circle of known center, orientation and radius~\footnote{The nearest distance $\Delta_*$ of a point ${\bf x_0}$ to a circle of radius $p$, in a plane normal to ${\bf N}$ centered at ${\bf x}_c$ is given simply by $\Delta_*^2 =  \Delta_\parallel^2+ (p-\Delta_\perp)^2$, where $\Delta_\parallel = (\xv_0-\xv_c) \cdot {\bf N}$ and $\Delta_\perp^2= |\xv_0-\xv_c|^2-\Delta_\parallel^2$.} greatly simplify calculations of inter-filament distances in a projected Hopf packing.  In $R^3$, these circular filaments have radius $p (\Phi)= \Omega^{-1} \sec \Theta$, they lie in planes tilted (transverse to the radial direction extending from $x=y=0$ axis) by $\Theta$ relative to $\hat{z}$ and their centers sit at ${\bf x}_c  (\Phi)= \rho (\Phi)(\sin \phi_0 \hat{x} - \cos \phi_0 \hat{y})$ such that they conform to the fixed-$\Theta$ tori.   Using this geometry to compute the distance $\Delta_{ij}(s_i)$ between the $i$th filament at $s_i$ along its length and the $j$th filament in terms of given coordinates $(\Theta_i, \phi_i)$ and $(\Theta_j, \phi_j)$ we consider a simple ``elastic" model for the cost to inter-filament cohesion due to inter filament strain,
\begin{equation}
\label{eq: stereoE}
E = \frac{1}{2} \sum_i \sum_{j\in \langle ij \rangle} \int ds_i |\Delta_{ij}(s_i)-d|^2 ,
\end{equation}
where the second sum runs over the nearest neighbors in a given packing to $i$ and $d$ is the filament diameter.  As a proxy for the optimal packings of $N$ cohesive discs on $S^2$, we take the positions of {\it icosadeltahedral tesselations} of sphere~\cite{siber}.  These tesselations, familiar to structural models of spherical viruses~\cite{caspar} and fullerenes~\cite{kroto}, are constructed from triangular tilings of icosohedra projected normally onto $S^2$ and are parameterized by the integer pair $\{m,n\}$ which describe the vector on separating centers of 5-fold coordination~\cite{siber}.

Figure \ref{fig: stereotori}B shows the packing energy per unit length $E/L$ of stereographically-projected Hopf bundles possessing icosadeltahedral order, where $L = \sum_i \int ds_i$ is the total Euclidean length of filaments in the bundle.  The strain energy density is plotted versus twist angle $\theta = \Theta$ of outer filaments for icosodeltahedral tesselations, for a range of tesselations from small twist, $\Omega_{\{4,2\}} = 0.0895 d^{-1}$, to large twist, $\Omega_{\{1,1\}} = 0.325 d^{-1}$.  Notably, the strain energy falls to zero as $\theta \to 0$, when the width of the bundles is small compared with the radius circular backbone, owing to the small conformal distortion near the $\Phi = 0$ ``pole" of the projection.  The characteristic increase in strain energy with twist in this case is {\it not} a symptom of the imperfect packing topology of filaments, as all $\theta\to \pi/2$ packings possess the topologically appropriate twelve five-coordinated sites needed for tessellations of $S^2$.  Rather, the increase in strain with $\theta$ in these projected Hopf packings is a reflection of the fact that filament spacings in the projection become locally over-(under-)dense on the inside (outside) of toroidal packing, and that conformal strain grows with toroidal thickness, roughly as $1-\omega \sim\sin \theta$ for small $\theta$.  The extent to which it is possible to relax the build up of strain energy at large $\theta$ via smooth relaxations of filament positions in projected packings, or if instead, the topological framework of the projected packings of $S^3$ are wholly inadequate candidates of optimal, large-$\theta$ bundles in $R^3$,  remains an open question.

\section{Concluding Remarks}

In conclusion, we have presented an emerging theoretical perspective on the unique metric geometry of complex, multi-filament or multi-column assemblies.  These studies show a powerful connection between the geometry of inter-filament spacing and the metric geometry of non-Euclidean surfaces.  The relationship between packing problems in filamentous assemblies with ``incompatible" textures and packing problems on intrinsically curved surfaces is particularly valuable  because physical models of optimal structure in the latter class of problems are well established, and the coupling between Gaussian curvature and topological defects in 2D membranes has received wide study in recent decades~\cite{giomi_bowick_advphys}.  Drawing on these familiar analogs sheds new light on the surprising rich, and  largely overlooked, questions of optimal structure in filamentous and columnar matter.  Furthermore, the purely geometrical origin of the frustration between patterns of orientation and spacing leads to a rich set of non-trivial and universal predictions for long-range order in a broad class of materials.  In particular, the optimal topological charge of the twisted packing was shown to be a universal function of a single geometric parameter, $\theta$ the tilt angle at the bundle surface, remarkably independent of elementary microscopic properties like filament interactions or diameter.   Considerations of inter-filament metric geometry are broadly applicable across material systems and material scales, and we anticipate, therefore, that the robust and geometrical origin of these predictions will aid in their direct experimental test.

The expanding resolution range offered by state of the art microscopy techniques down to nanometer and sub-nanometer scale, provides an exciting opportunity to test universal predictions for geometrically frustrated fibers and poses important, new challenges for their theoretical understanding. For example, recent high-resolution cryoTEM studies of DNA confined within bacteriophage capsids by Livolant and Leforestier reveal a surprisingly detailed picture of inter-strand organization taking place within what is clearly a highly frustrated and heterogeneous packing~\cite{livolant_pnas_09}.   DNA chains exhibit the seemingly contradictory combination of locally six-fold (hexagonal) packing, high degree of order and density throughout the roughly spherical volume.  While current imaging achieves sub-strand resolution only within transverse 2D sections, full three-dimensional reconstruction of the positions and orientations through such a complex packing may soon be achievable.  

Understanding interplay between the  texture induced by spherical confinement and the complex spectrum of topological defects in the transverse  packing in a maximally dense assemblies, what might be viewed as the filamentous analog to the Thomson problem, introduces several key challenges.  Specifically, how are constraints of inter-filament metric geometry formulated under conditions where the texture itself varies throughout?  Surely, a fully covariant formulation of the elasticity of columnar structures would provide a valuable tool for tackling optimal structure where assumptions about small tilt relative to a well-defined (and effectively Euclidean) reference state cannot be maintained.  While no fundamental obstacles prevent the development of such a covariant theory, it remains to be seen how well such a theory may illuminate properties of optimal packing where inter-filament metric geometry cannot be reduced to a single curved 2D manifold, but may instead require packing to be optimized over a sequence of inequivalent surfaces representing variation of inter-filament texture throughout a structure as complex as a confined, contorted and folded chain packings.

\begin{acknowledgments}
I am grateful to collaborators I. Bruss and A. Azadi for numerous discussions and insights on topics reviewed in the article.   Further, I thank L. Cajamarca, J.-F. Sadoc and P. Ziherl for critical readings of this manuscript, and O. Lavrentovich for discussions valuable for stimulating this review.  This work was supported by NSF CAREER Award DMR 09-55760 and through an award from the Alfred P. Sloan Foundation.\end{acknowledgments}

\bibliography{bib_v2.bib}

\end{document}